\PassOptionsToPackage{table}{xcolor}

\documentclass[sigconf, nonacm]{acmart}

\usepackage{tikz}
\usepackage{amsmath}
\usepackage{amsmath,amsfonts,amsthm}
\usepackage{algorithmic, algorithm}
\usepackage{graphicx}
\usepackage{textcomp}
\usepackage{xcolor}
\usepackage{colortbl}
\usepackage[inline, shortlabels]{enumitem}
\usepackage{adjustbox}
\usepackage{multirow}
\usepackage{threeparttable}
\usepackage{subfigure}
\usepackage{mathtools}
\usepackage{makecell}
\usepackage{bm}

\newcommand\vldbdoi{XX.XX/XXX.XX}
\newcommand\vldbpages{XXX-XXX}
\newcommand\vldbvolume{16}
\newcommand\vldbissue{1}
\newcommand\vldbyear{2022}
\newcommand\vldbauthors{\authors}
\newcommand\vldbtitle{\shorttitle} 
\newcommand\vldbavailabilityurl{https://github.com/ANL-FAIR-SBI/SOLAR}
\newcommand\vldbpagestyle{plain}

\newcommand{\del}[1]{}

\newcommand{\TheWork}{\textsc{SOLAR}}
\newcommand{\TableStyle}{\centering\footnotesize\sffamily}

\newcommand{\newstuff}[1]{{\color{blue}\sffamily #1}}
\newif\iffinal

\iffinal
    \newcommand\xiaodong[1]{}
\else
    \newcommand\xiaodong[1]{{\color{red}[Xiaodong: #1]}}
\fi

\begin{document}

\title{SOLAR: A Highly Optimized Data Loading Framework for Distributed Training of CNN-based Scientific Surrogates}
    
\settopmatter{authorsperrow=5}

\newcommand{\iu}{Indiana University}
\newcommand{\anl}{Argonne National Lab}

\newcommand{\AFFIL}[3]{%
    \affiliation{%
        \institution{\small #1}
        \city{\small #2}\state{\small #3}\country{\small USA}
    }
    }

\newcommand{\IU}{\AFFIL{\iu}{Bloomington}{IN}}
\newcommand{\ANL}{\AFFIL{\anl}{Lemont}{IL}}

\renewcommand{\thefootnote}{\fnsymbol{footnote}}

\author{Baixi Sun}{\IU}
\email{sunbaix@iu.edu}

\author{Xiaodong Yu\footnotemark[1]{}}{\ANL}
\email{xyu@anl.gov}

\author{Chengming Zhang}{\IU}
\email{czh5@iu.edu}

\author{Jiannan Tian}{\IU}
\email{jti1@iu.edu}

\author{Sian Jin}{\IU}
\email{sianjin@iu.edu}

\author{Kamil Iskra}{\ANL}
\email{iskra@mcs.anl.gov}

\author{Tao Zhou}{\ANL}
\email{tzhou@anl.gov}

\author{Tekin Bicer}{\ANL}
\email{tbicer@anl.gov}

\author{Pete Beckman}{\ANL}
\email{beckman@anl.gov}

\author{Dingwen Tao\footnotemark[1]{} \footnotetext[1]{Authors to whom correspondence should be addressed: Xiaodong Yu (xyu@anl.gov) and Dingwen Tao (ditao@iu.edu).}}{\IU}
\email{ditao@iu.edu}

\renewcommand{\shortauthors}{Sun et~al.}

\begin{abstract}

CNN-based surrogates have become prevalent in scientific applications to replace conventional time-consuming physical approaches. Although these surrogates can yield satisfactory results with  significantly lower computation costs over small training datasets, our benchmarking results show that data-loading overhead becomes the major performance bottleneck when training surrogates with  large datasets. In practice, surrogates are usually trained with high-resolution scientific data, which can easily reach the terabyte scale. Several state-of-the-art data loaders are proposed to improve the loading throughput in general CNN training; however, they are sub-optimal when applied to the surrogate training. In this work, we propose \TheWork{}, a  \textbf{\underline{s}}urr\textbf{\underline{o}}gate data \textbf{\underline{l}}o\textbf{\underline{a}}de\textbf{\underline{r}}, that can ultimately increase loading throughput during the training. It leverages our three key observations during the benchmarking and contains three novel designs. Specifically, \TheWork{} first generates a pre-determined shuffled index list and accordingly optimizes the global access order and the buffer eviction scheme to maximize the data reuse and the buffer hit rate. It then proposes a tradeoff between lightweight computational imbalance and heavyweight loading workload imbalance to speed up the overall training. It finally optimizes its data access pattern with HDF5 to achieve a better parallel I/O throughput. Our evaluation with three scientific surrogates and 32 GPUs illustrates that \TheWork{} can achieve up to 24.4$\times$ speedup over PyTorch Data Loader and 3.52$\times$ speedup over state-of-the-art data loaders.
\end{abstract}

\maketitle

\setcounter{tocdepth}{4}

\pagestyle{\vldbpagestyle}
\begingroup\small\noindent\raggedright\textbf{PVLDB Reference Format:}\\
\vldbauthors. \vldbtitle. PVLDB, \vldbvolume(\vldbissue): \vldbpages, \vldbyear.\\
\href{https://doi.org/\vldbdoi}{doi:\vldbdoi}
\endgroup
\begingroup
\renewcommand\thefootnote{}\footnote{\noindent
This work is licensed under the Creative Commons BY-NC-ND 4.0 International License. Visit \url{https://creativecommons.org/licenses/by-nc-nd/4.0/} to view a copy of this license. For any use beyond those covered by this license, obtain permission by emailing \href{mailto:info@vldb.org}{info@vldb.org}. Copyright is held by the owner/author(s). Publication rights licensed to the VLDB Endowment. \\
\raggedright Proceedings of the VLDB Endowment, Vol. \vldbvolume, No. \vldbissue\ %
ISSN 2150-8097. \\
\href{https://doi.org/\vldbdoi}{doi:\vldbdoi} \\
}\addtocounter{footnote}{-1}\endgroup
\renewcommand{\thefootnote}{\arabic{footnote}}

\ifdefempty{\vldbavailabilityurl}{}{
\vspace{.3cm}
\begingroup\small\noindent\raggedright\textbf{PVLDB Artifact Availability:}\\
The source code, data, and/or other artifacts have been made available at \url{\vldbavailabilityurl} (per request by email).
\endgroup
}

\section{Introduction}\label{sec:introduction}


\textbf{CNN Surrogates}
Deep neural networks (DNNs) such as convolutional neural networks (CNNs) are critical to traditional artificial intelligence (AI) tasks such as image and vision recognition \cite{he2016deep,voulodimos2018deep}, recommender systems \cite{he2020lightgcn,zhang2019deep}, and natural language processing \cite{collins2001convolution,alshemali2020improving}.
Very recently, domain scientists such as physicists and chemists also leverage neural network techniques to build surrogates for their scientific applications to substitute conventional physical solutions~\cite{obiols2020cfdnet, dong2020smart, dong2019adaptive, yin2022strategies, shi2022gnn}.
Conventional solutions built upon physical principles usually consist of iterative processes with complex operations.
They can provide accurate results, however, are quite time-consuming, even with the accelerations
on high-performance computing (HPC) platforms. For instance, the traditional iterative phase retrieval 
algorithm for ptychographic image reconstruction takes hours or even days to converge on multiple GPUs
with only a medium-scale dataset~\cite{yu2022scalable}.
To accelerate the reconstruction, physicists develop a CNN-based autoencoder (named PtychoNN \cite{PtychoNN}) to approximate the reconstructed images from raw X-ray data and achieve 300$\times$ speedup compared to the traditional physical approach~\cite{cherukara2020ai}.
Another example is CosmoFlow~\cite{mathuriya2018cosmoflow}, which is a CNN-based model used to predict the cosmological parameters in the universe and is more than 6,000$\times$ faster than the traditional physical approach when processing large 3D dark matter parameters on a single compute node.
All these models, called \textit{CNN-based scientific surrogate models},\!\footnote{We interchangeably use (CNN) surrogate models and surrogates in this paper.} allow scientists to quickly get acceptable approximate results without performing traditional time-consuming physical algorithms.

\textbf{Performance Bottleneck}
Although CNN surrogates have been approved to be efficient with small datasets, their training performances with 
large-scale data on the HPC platforms remain unclear. To obtain insights, we first benchmark the training of some 
representative surrogates. Based on the benchmarking results (see~\S\ref{sec:mot}),
we observe that, in contrast to the training in conventional AI tasks, \textit{data loading} is the major bottleneck 
of surrogate training (occupies 98\% of training time in some cases). There are two possible reasons to justify this observation:
\begin{enumerate*}
    \item The networks in surrogate models are relatively simple and have a small number of parameters. For example, PtychoNN~\cite{cherukara2020ai} consists of only 1.2 million parameters.
    \item The training sets of surrogates usually are high-resolution scientific data~\cite{kurth2018exascale, mathuriya2018cosmoflow}, and thus the total dataset sizes are immense. For instance, the training dataset
    of CosmoFlow~\cite{mathuriya2018cosmoflow} is over 10 TB consisting of $512^3$ high-resolution 3D samples~\cite{oyama2020case}.
\end{enumerate*}
As a comparison, ResNet-50, a classical network for image recognition, has 23 million parameters and is typically trained
only on a 150GB dataset (e.g., ImageNet \cite{deng2009imagenet}). 
Accordingly, while the computations for updating networks 
dominate the training performance of classical AI tasks, the top challenge in surrogate training is to 
efficiently manage and load the huge scientific training dataset \cite{zhang2021distributed}.

\textbf{Research Challenges}
A common approach for reducing the data loading cost is to buffer the dataset in the near-processor 
memories, so that loading data from remote file systems (e.g., PFS) becomes a one-time effort.
However, in many cases, especially with large-scale training datasets, the memory space is not 
enough to buffer the entire dataset, and buffer misses will happen. Because of the random data 
shuffling, 
the buffer misses are actually frequent when loading each local mini-batch.
Therefore, we still need to load samples from either PFS or neighbor nodes (in distributed training)
in each training step. Although the buffering scheme can alleviate the data loading overhead, 
it causes three issues:
\begin{enumerate*}[label=(\roman*)]
    \item The buffer hit rate cannot be guaranteed due to data shuffling in each step.
    \item The loading workload for each device can be severely imbalanced.
    \item It leads to frequent random accesses to small data fragments, making the parallel I/O libraries (e.g., HDF5 \cite{hdf5}) inefficient.
\end{enumerate*}

\textbf{Existing Works}
There are a few state-of-the-art data loaders have been proposed to tackle some of the 
above issues caused by the buffering scheme.
\begin{enumerate*}[label=(\arabic*)]
    \item NoPFS~\cite{dryden2021clairvoyant} utilizes a heuristic performance model to predict the data to be used by the next epoch, and accordingly determine the data eviction scheme to increase the buffer hit rate.
    \item DeepIO~\cite{zhu2018entropy} limits the data shuffling within the buffer of each compute node to eliminate the buffer misses and achieve maximum data reuse.
    \item Lobster~\cite{liu2022lobster} improves NoPFS and balances the loading workload by giving different numbers of loading processes to different devices.
\end{enumerate*}
Although these data loaders can achieve improved performance in 
general training tasks, they are sub-optimal when applied to surrogate training with large
scientific datasets. 
\begin{enumerate*}[label=(\arabic*)]
    \item NoPFS can only approximately optimize the prefetching for only the next epoch, and thus can not maximize the data reuse in most cases.
    \item DeepIO significantly reduces the randomness and accordingly will sharply degrade the training accuracy of surrogates, which is not acceptable as the outputs of scientific surrogates usually are high-resolution.
    \item Lobster trades the loading process imbalance for a balanced loading workload. This tradeoff is less efficient in surrogate training, as data loading takes the majority time of the entire training, hence the process imbalance exists during most training time, and its overhead is nontrivial. 
\end{enumerate*}
Moreover, there is no existing work addressing the third issue.
\textbf{Our Contributions}
To ultimately mitigate the aforementioned data loading issues in surrogate training, in this work, 
we design a data loading framework, called \TheWork{}, that can drastically improve the loading efficiency
during the distributed training of CNN surrogates. Our design is fundamentally different from the 
state-of-the-art data loaders (detailed differences will be elaborated in~\S\ref{sec:design})
and is based on our key observations during surrogate training:
\begin{enumerate*}[label=(\roman*)]
    \item The shuffled sample indices for all epochs can be determined prior to the training, and reordering the sample indices within the global batch makes no change to the randomness and keeps the same training accuracy.
    \item The computation imbalance is much less harmful than the data loading workload imbalance during the surrogate training.
    \item A chunked large data load can be more efficient than multiple random small data loads with HDF5, even if the overall loaded size is bigger.
\end{enumerate*}
Accordingly, we propose three novel designs in our framework:
\begin{enumerate*}[label=(\roman*)]
    \item We avoid the runtime shuffling after each epoch and generate a complete shuffled index list for all epochs ahead of training. This pre-determined list enables two orthogonal offline optimizations: (1) re-arranging the epoch order to optimize the buffer eviction scheme, and (2) changing the mapping from data samples to devices to increase the data locality. Compared to NoPFS, this design leverages fully determined sample indices and can maximize data reuse globally.
    \item We trade the computation imbalance for a balanced data-loading workload. Different from Lobster, our tradeoff cost is negligible since computation is only a tiny portion of the surrogate training.
    \item We aggregate multiple single sample accesses within a certain locality into a chunk load to improve the parallel HDF5 loading throughput. 
\end{enumerate*}
The contributions of our entire work are summarized as follows.
\begin{itemize}[noitemsep, topsep=0pt, leftmargin=1.3em]
    \item We conduct performance benchmarking on three surrogate models (PtychoNN, AutoPhaseNN, and CosmoFlow) and identify their major performance issues during the training. 
    \item We propose \TheWork{}, a data loading framework using fundamentally different designs compared to the existing works and can ultimately mitigate the surrogate training issues.
    \item We generate the pre-determined shuffled index list in \TheWork{} and accordingly introduce two optimizations. This design allows \TheWork{} to change the epoch order to optimize the buffer eviction scheme globally and maximize the buffer hit rate and data reuse.
    \item We propose a tradeoff between lightweight computational imbalance and heavyweight data-loading imbalance in \TheWork{} to improve the overall training performance.  
    \item We optimize the data access pattern with HDF5 to maximize \TheWork{}'s parallel I/O throughput. 
    \item We evaluate \TheWork{} using three surrogates over five datasets with various system configurations. The results demonstrate that \TheWork{} achieves up to 24.43$\times$ reduction in data-loading time and 3.03$\times$ speedup in overall training performance.
\end{itemize}

The remaining paper is organized as follows. 
First, we discuss the research background in \S\ref{sec:background} and motivation in \S\ref{sec:mot}.
Second, we present our design of \TheWork{} in \S\ref{sec:design} and our evaluation results in \S\ref{sec:evaluation}.
Third, we discuss related work in \S\ref{sec:related}.
Finally, we conclude our work and discuss future plans in \S\ref{sec:conclusion}.
\section{Background}
\label{sec:background}


\subsection{DNN-Based Surrogates}\label{sec:dnn_surro}
\begin{figure}
    \centering
    \includegraphics[width=\linewidth]{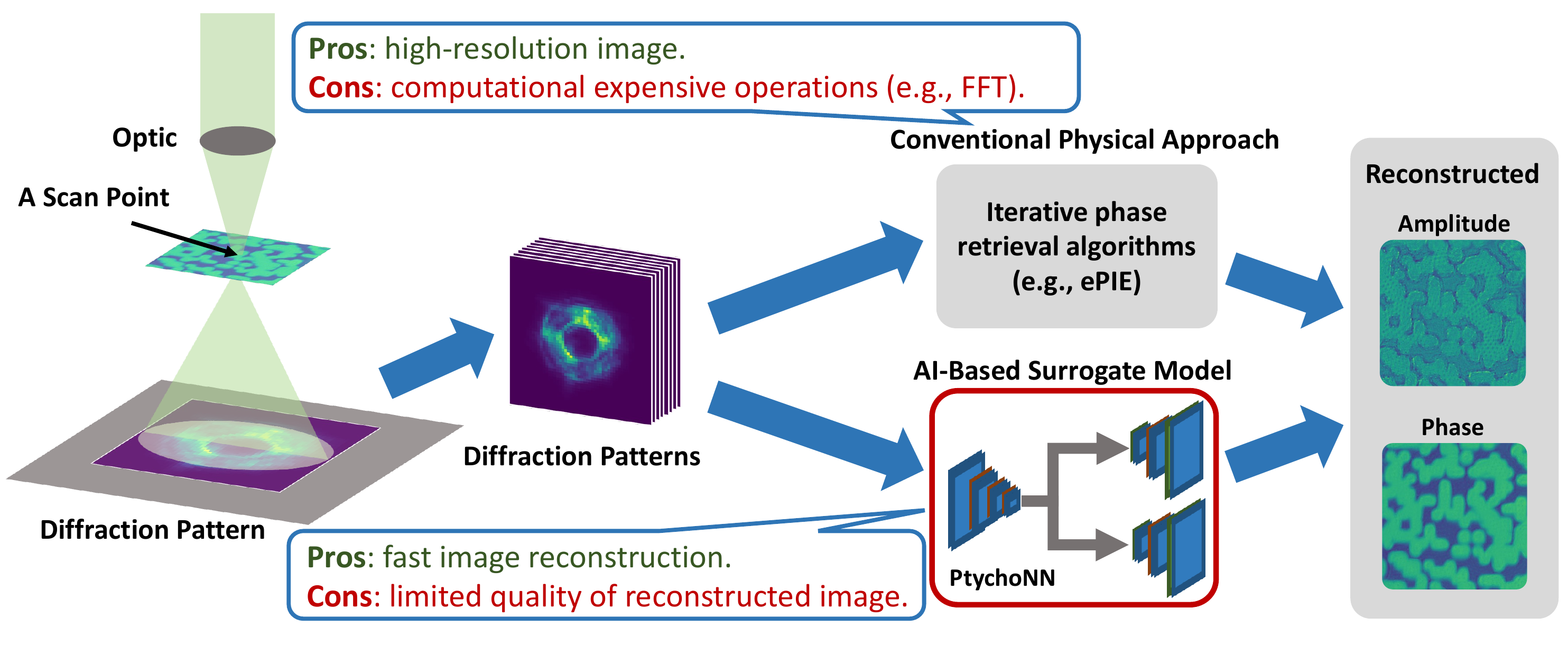}
    \caption{Workflow of ptychographic image reconstruction. 
    The CNN-based surrogate model achieves hundreds of times speedup over iterative phase retrieval algorithms~\cite{cherukara2020ai}.
    }
    \label{fig:pty_vs}
    \vspace{-2mm}
\end{figure}

DNN-based surrogate models are designed to substitute heavy-cost traditional iterative algorithms in scientific simulations. For example, Figure~\ref{fig:pty_vs} shows the workflow of reconstructing images from scanned diffraction patterns.
Specifically, AI-based surrogate models achieve 300$\times$ speedup in the image reconstruction task compared to the conventional physical approaches, with minor loss of the image resolution.

However, the problem size is becoming huge due to the growing size of scientific datasets.
For example, the coherent imaging data can be generated in terabytes per hour by the Advanced Photon Source Upgrade (APS-U) \cite{noirot2018workshop}.
When the dataset size cannot fit into the memory of a single device (i.e., CPU, GPU), a straightforward solution is to use mini-batch training, which loads a small batch of data at a time for each step. Since the training time would be long with a single device,
it often requires parallelizing the workload to multiple devices using data parallelism, especially for training DNN-based surrogates with large-scale input data. 

\subsection{Distributed Training with Data Parallelism}
Data parallelism techniques are widely utilized for training on large datasets~\cite{mathuriya2018cosmoflow, kurth2018exascale, deng2009imagenet, farrell2021mlperf}.
When applying data parallelism on HPC clusters, the training typically includes three stages: (1) 
I/O: loading the data from a remote parallel file system (i.e., GPFS, Lustre) to host memory; (2) Computation: performing forward and backward phases to calculate the local gradient on each device; (3) Communication: synchronizing averaged gradients across multiple devices to update model weights. Among these stages, I/O is a bottleneck for distributed training with large-scale datasets~\cite{dryden2021clairvoyant,pumma2019scalable, zhu2018entropy}. 
Thus, \textit{\TheWork{} is designed primarily to optimize the data loading stage in training DNN-based surrogates}.
The access order is crucial for data reuse in data loading \cite{pumma2017towards}.
In most DNN training, each data sample is accessed only once during an epoch.
Thus, to reuse the data, the buffer needs to keep the samples that will be used soon and evict those that will be used further in the future~\cite{dryden2021clairvoyant}.
However, random shuffling makes it hard to achieve a good reuse rate and reduce data loading time.
Existing approaches such as DeepIO~\cite{zhu2018entropy} change the access order to achieve a better data reuse rate.
However, these methods reduce the randomization of the dataset and degrade the training accuracy. On the other hand, an existing work ~\cite{yang2019accelerating} proved
that by adjusting the access order in a particular way, the synchronized gradients will be the same as the original random shuffle.
Specifically, it proves that reordering the data samples within the global batch produces the same gradients after synchronization.  
This inspires us to create a better data reuse method without affecting the training accuracy.
\subsection{I/O Frameworks for Distributed Training}
DataLoader in PyTorch~\cite{paszke2019PyTorch} is the default data loading tool and for data parallelism. It has two optimizations: 
(1) using a subprocess to read data in order to fully utilize the bandwidth; 
(2) prefetching the training data to overlap data loading and computation. 
However, when the data loading dominates the epoch, GPUs are idle and starve due to a lack of data. 
To this end, some works store the data that are accessed in the current epoch to reuse it in the next epoch.

Dryden \textit{et al.}~\cite{dryden2021clairvoyant} proposed NoPFS, a prefetching and buffering strategy to reduce the data loading time by reusing the accessed data samples.
The optimization is twofold: (1) combining the clairvoyant method with the distribution of samples on multiple nodes to optimize data eviction policy and (2) designing a performance model to leverage the multi-layer architecture of the HPC cluster.
Liu \textit{et al.} proposed Lobster~\cite{liu2022lobster}, a dynamic thread arrangement and data eviction strategy that further improves the data-loading performance by balancing the data-loading process across multiple GPUs.
Their optimizations are twofold: (1) rearranging the threads to read and preprocess the data samples to balance the data loading, and (2) tuning the data sample eviction to achieve better data reuse. 
However, both methods focus only on the online buffer eviction strategy and do not obtain the same observations as \TheWork{}; thus, their performance improvements in data loading are limited for training surrogate models.

\section{Research Motivation}\label{sec:mot}
In this section, we present the benchmarking results of the distributed surrogate training.
We first illustrate the scalability of training a typical CNN-based surrogate with three DL frameworks.
We then break down the training time and exhibit that data loading is the major bottleneck taking up 
to 98.6\% of the training time. These results motivate us to explore more efficient data-loading
solutions for surrogate training.

We conduct the benchmarking on ThetaGPU~\cite{thetaGPU} (each node has eight A100 GPUs) at the Argonne Leadership Computing Facility, following the guidelines from MLPerf HPC~\cite{farrell2021mlperf}. 

\begin{figure}[t]
    \centering
    \includegraphics[width=0.85\linewidth]{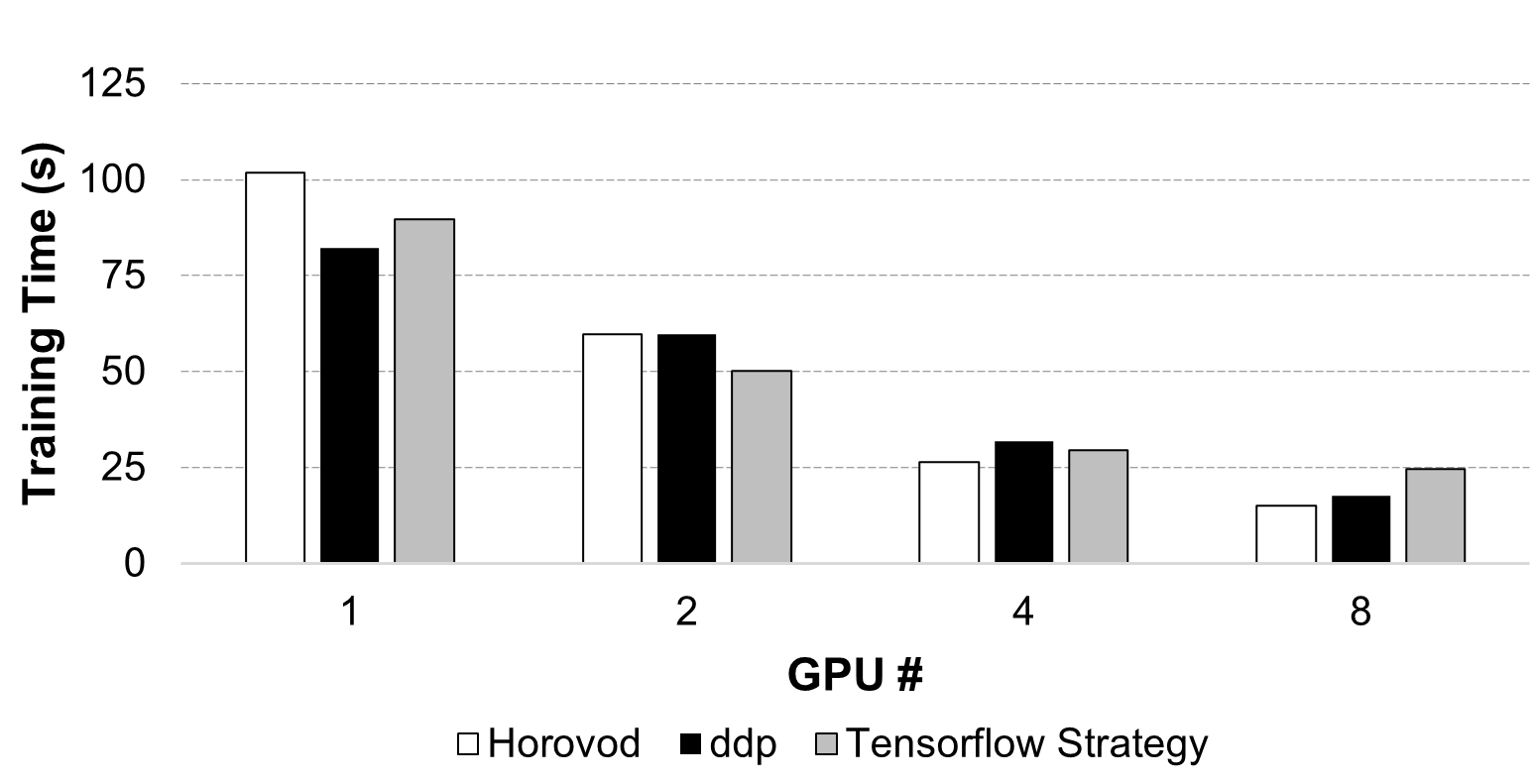}
    \caption{
    Training time in seconds using different numbers of GPUs with different frameworks.
    }
    \label{fig:bench_scale}
    \vspace{-2mm}
\end{figure}

\vspace{3pt}
\textbf{Scalability Study}
We first perform a study to show the scalability of distributed training over surrogate models with widely used deep learning frameworks.
Specifically, we 
implement PtychoNN with TensorFlow mirrored strategy~\cite{abadi2016TensorFlow}, Horovod~\cite{sergeev2018horovod}, and PyTorch DDP \cite{ddp}.
We train PtychoNN on a 17 GB dataset (will be detailed in \S \ref{sec:evaluation}) with different numbers of GPUs.
Specifically, we take the average epoch time among 99 epochs, excluding the  warmup time in the first epoch.
Figure~\ref{fig:bench_scale} shows that these three frameworks
can provide similar scalability from a single to 8 GPUs. Therefore, we choose Pytorch DDP as the distributed computation tool for the experiments in the rest of this paper.

\begin{figure}
    \centering
    \includegraphics[width=\linewidth]{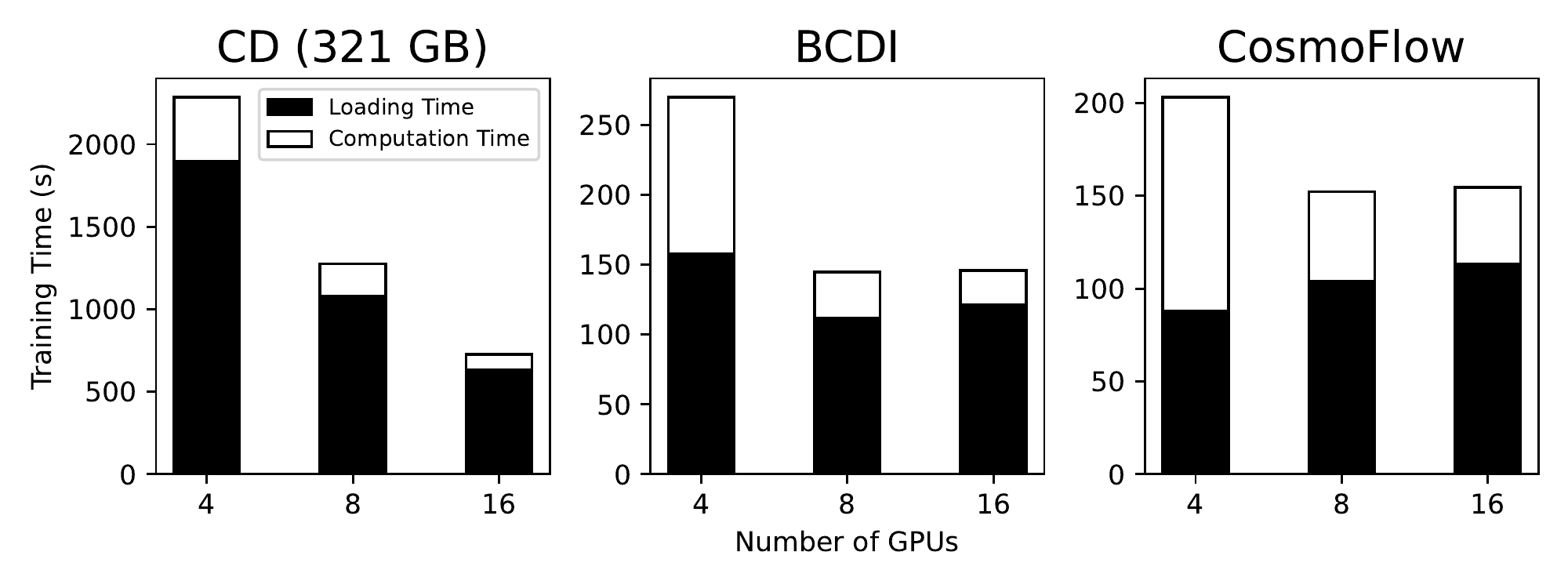}
    \caption{Time breakdown on tested datasets when using prefetch to overlap data loading with computation. 
    }
    \vspace{-2mm}
    \label{fig:321break}
\end{figure}

\begin{table}[t]
    \TableStyle
    \caption{Breakdown of computation time and data loading time (in seconds) when training PtychoNN on 1.2 TB dataset.}
    \resizebox{\linewidth}{!}%
    {
        \begin{tabular}{@{} >{\bfseries}l rrrrrr @{}}
            \toprule
            {\#GPU}        & \multicolumn{2}{c}{32} & \multicolumn{2}{c}{64} & \multicolumn{2}{c}{128}                                        \\
            \cmidrule(lr){2-3}\cmidrule(lr){4-5}\cmidrule(l){6-7}
            {data loading} & 307.7             & 1.00$\times$          & 159.7                 & 1.93$\times$ & 80.2  & 3.83$\times$ \\
                           & 98.5\%                &                       & 98.6\%                 &              & 98.6\% &              \\
            \cmidrule(lr){2-3}\cmidrule(lr){4-5}\cmidrule(l){6-7}
            {computation}  & 4.7                & 1.00$\times$          & 2.3                 & 2$\times$ &  1.1  & 4.13$\times$ \\
                           & 1.5\%                &                       & 1.4\%                 &              & 1.4\% &              \\
            \midrule
            total          & 312.3                & 1.00$\times$          & 162.0                 & 1.93$\times$ & 81.4  & 3.84$\times$ \\
            \bottomrule
        \end{tabular}
    }
    \label{tab:my_label}
\end{table}
\vspace{3pt}
\textbf{Performance Breakdown}
We then break down the overall performance to investigate the bottlenecks in distributed training of surrogate models.
We train three surrogate models on the corresponding datasets, i.e., pythcoNN \cite{cherukara2020ai} on the Coherent Diffraction (CD) 321GB dataset and 1.2TB dataset, AutoPhaseNN \cite{yao2022autophasenn} on the 151GB Bragg Coherent Diffraction Imaging (BCDI) dataset, and CosmoFlow \cite{mathuriya2018cosmoflow} on a 1TB subset of the 3D universe simulation (CosmoFlow) dataset (datasets will be detailed in \S \ref{sec:evaluation}).

The breakdown times are shown in Figure~\ref{fig:321break} and Table~\ref{tab:my_label}.
Specifically, Figure~\ref{fig:321break} shows the time breakdown to train three surrogates on three datasets using different numbers of GPUs.
We also synthesize a 1.2TB dataset for ptychoNN to indicate how different sizes of training datasets impact the training performance of the same surrogate model, as shown in Table~\ref{tab:my_label}.

In Figure~\ref{fig:321break} we can observe that the data loading is the major performance bottleneck.
Specifically, when using four GPUs, data loading takes 83.1\%, 77.3\%, and 43.2\% of the total training time with PtychoNN, AutoPhaseNN, and CosmoFlow, respectively.
Along with the increase of the GPU number (i.e., a weak scaling), the computation time reduces quickly whereas the data loading cost drops very slowly (even increasing in some cases).
Consequently, the data-loading issue becomes more severe when using more computation resources.
For instance, When we increase the GPU number from four to 16 for CosmoFlow training with the 3D universe simulation dataset, 
the percentage of data loading time increases from 43.2\% to 73.4\% of the total training time.
In addition, in Table~\ref{tab:my_label}, we can observe that the data-loading problem also 
can become worse when the dataset gets larger, as it takes up to 98.6\% of the total training time when 
training ptychoNN on the 1.2TB dataset. We note that we use more GPUs for training with the 1.2TB data 
in order to keep a reasonable computation workload on each GPU.

Moreover, we observe that our achieved data-loading throughput with parallel HDF5 is lower than that of reported by the HDF5 Group~\cite{byna2020exahdf5}. 
This is because random accesses with small data degrade the data-loading efficiency, which will be discussed in detail in \S\ref{sec:aggregated}.

\vspace{3pt}
\textbf{Motivation}
The above results and observations motivate us to design an efficient framework to optimize data loading time for CNN-based surrogates by developing offline scheduling and runtime buffering strategies.
\section{Proposed Design of SOLAR}
\label{sec:design}


\subsection{Overview}
\begin{figure*}[ht!]
    \centering
    \includegraphics[width=\linewidth]{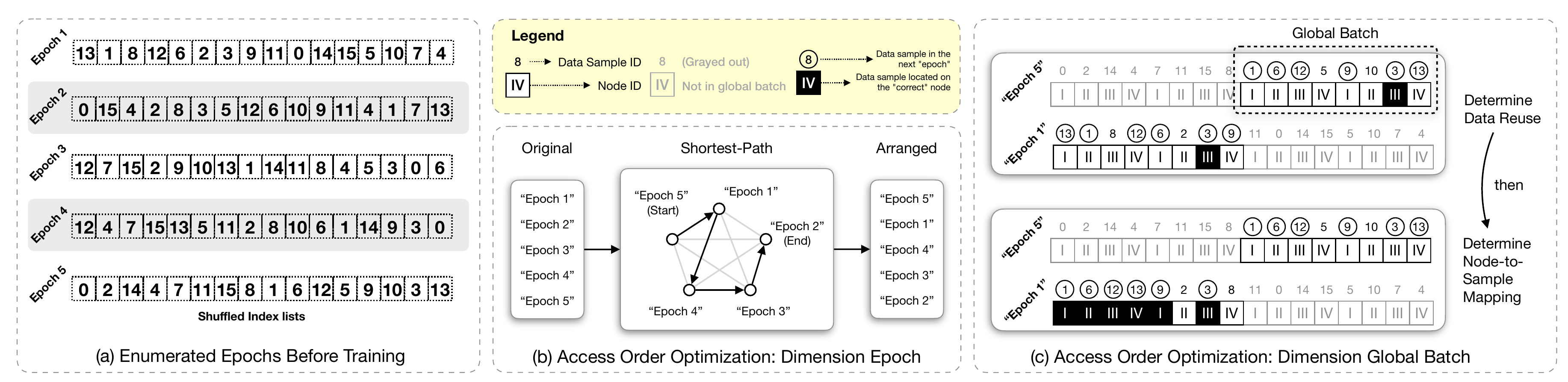}
    \caption{Our proposed offline scheduling: (a) Obtain access order based on random shuffle before training;
    (b) Adjust epoch order  by solving a graph problem to maximize data reuse;
    (c) Re-arrange node-to-sample mapping to optimize data locality.}
    \vspace{-2mm}
    \label{fig:overview}
\end{figure*}
To reduce the data-loading overhead, a common solution is to buffer the loaded samples in a faster memory space 
(e.g., DRAM) for reuses in the following epochs. The buffering approach theoretically can reduce the number of 
samples to be re-loaded from PFS. However, it usually cannot reach the expected throughput improvement due to 
the following issues:
\begin{enumerate*}
    \item The buffer hit rate could be low due to the sample shuffling after each epoch.
    \item The loading workload of each device is imbalanced because of the buffer hit rate variability on different devices.
    \item Reloading evicted samples from PFS becomes random access to small fragmented data, which aligns poorly with the desired access pattern of parallel HDF5.  
\end{enumerate*}

Our \TheWork{} framework addresses the above issues based on our key observations during the surrogate training:
\begin{enumerate*}[label=(\arabic*)]
    \item We can predetermine the shuffled sample indices for all epochs and reorder the indices within a global batch, without affecting the training accuracy.
    \item The impact of computation imbalance is trivial compared to that of data-loading workload imbalance.
    \item Loading a chunked large data load can be faster than loading several small data fragments with HDF5, even though the chunked one loads redundant data.
\end{enumerate*}
Accordingly, \TheWork{} proposes three novel designs (one for each issue):
\begin{enumerate*}[label=(\arabic*)]
    \item It generates a shuffled index list prior to the training, and then re-arranges the epoch order and changes the sample-device mapping according to the list to maximize the data reuse.
    \item It trades the computation imbalance for a balanced data-loading workload for better overall training performance.
    \item It uses a large chunked loading to replace small fragmented loadings within a certain locality to improve the loading throughput.
\end{enumerate*}


We implement these designs in \TheWork{} as an offline scheduler and a runtime buffering strategy, which will be demonstrated in Figure~\ref{fig:overview} and Figure~\ref{fig:runtime}.
The offline scheduler includes optimizing the access order and computing the temporal (when to load) and spatial (where the data is located) information about each data sample.
The runtime buffering strategy includes data eviction optimization and an optimized chunk-loading mechanism.
Also, the workflow and architecture for the runtime buffering strategy is illustrated in Figure~\ref{fig:runtime}. Table~\ref{table:variables} lists the notation used throughout this paper. To our best knowledge, \TheWork{}
is the first work obtaining the above observations and insights in surrogate training, and accordingly design optimizations for surrogate data loading.

\begin{table}[]
    \TableStyle
    \caption{Notation used in this paper.}
    \begin{tabular}{@{} l ll  @{}}
        \toprule
        \bfseries Variable   & \bfseries Unit & \bfseries Definition                                   \\
        \midrule
        $|\text{Buffer}|$    & image          & Buffer Size                                            \\ %
        $|\text{Batch}|_{l}$ & image          & Local Batch Size                                       \\
        $|\text{Batch}|_{g}$ & image          & Global Batch Size                                      \\
        $|\text{Dset}|$      & image          & Dataset Size                                      \\
        $|\text{chunk}|$     & image          & Optimal chunk size in aggregated chunk loading \\
        $G$                  &                & A graph in which each vertex represents an epoch       \\
        $p$                  &                & A path within graph $G$                                \\
        $N_{u,v}$            & image          & The number of samples to load from epoch $u$ to epoch $v$ \\
        $n$                  &                & A compute node on HPC cluster                          \\
        $E$                  &                & Number of epochs                                       \\
        $e_i$                &                & The $i$-th epoch                                       \\
        \bottomrule
    \end{tabular}
    \label{table:variables}
\end{table}
\subsection{Maximizing Data Reuse}
\subsubsection{Optimizing Epoch Order}\hfill\\
\textbf{Limitations of Existing Work}
NoPFS~\cite{dryden2021clairvoyant} tries to increase the data reuse by optimizing the buffer eviction scheme. 
However, since it generates the shuffle list at runtime after each epoch, it has to utilize the prediction of data sample distribution with a performance model to determine the runtime prefetching and caching strategy.
Meanwhile, it considers only the data reuse problem between each pair of epochs.
By doing so, the solution provided by NoPFS to optimize the data-loading performance is suboptimal.
In comparison, \TheWork{} generates the shuffle list for all epochs offline.
Specifically, our benefits are two-fold: (1) instead of considering only the data reuse in the next epoch, \TheWork{} generates the shuffle list for all epochs so that we can have a global view for optimization; (2) instead of determining when and where to fetch data  \textit{online} and in a complicated fashion, we perform \textit{offline} access order optimizations that maximize the data reuse rate.\!\footnote{Note that the access order optimization is a one-time overhead. 
}

\textbf{Observation}
To address the first issue, we leverage a key observation that instead of generating during runtime, the shuffled index list can be determined ahead of training. Since the data reuse rate varies when we change the order of epochs, we can optimize the epoch order according to the list to maximize the data reuse.
Here, the epoch order is defined as the order of using the shuffled index lists,
where each shuffled index list indicates the data samples' access order for an epoch.
In distributed training, each data sample is accessed only once in each epoch.
That is to say, we can only reuse the loaded data in the next epoch.
In practice, the result of the random shuffle (shuffled index list) is determined with a given random seed.
Furthermore, parameters like the number of epochs, number of devices, and random seed are pre-defined. 
Thus, instead of performing the random shuffle at the beginning of each epoch, we can produce the shuffled index list for all epochs before training and then design offline optimizations based on these shuffled index lists.
In terms of data reuse, the data samples must be loaded if they have not already been stored in the current process during the last epoch.
Similar to the ``cache hit,\!'' a high data reuse rate between two epochs is preferred to reduce the data loading operations.
Through experiments, we observe that the data reuse rate can vary with the epoch order.
Thus, we propose an algorithm to arrange the epoch order to maximize the data reuse.

\textbf{Proposed Optimization}
With the above important observation, we design an optimization that maximizes data reuse by adjusting the order of epochs.
First, we generate the shuffled index list for all epochs using the defined random seed, as shown in Figure~\ref{fig:overview} (a).
Instead of performing the shuffle operation for each epoch, the ahead-of-time random shuffle procedure is conducted offline to obtain the access order for all epochs before training.
By doing so, we are able to perform offline optimizations according to the access order.
Next, we define the problem as finding an epoch order such that the data reuse is maximized.
Then, we abstract this problem into a graph problem.
Consider the shuffle list as a graph $G$ whose nodes represent all epochs, and the edge weight denotes the loading cost from one epoch to the other; we are solving the problem by finding the shortest path that visits each node exactly once.
Let $N_{u,v}$ denote the number of data samples to load from epoch $u$ to epoch $v$; we compute the edge weight by
\begin{equation}\label{eq:model}
    N_{u,v} = \mathbf{card}(\text{Buffer}_v-\text{Buffer}_u),
\end{equation}
where $\textit{Buffer}_v$ means the set of data in epoch $v$'s \textit{last} buffer, $\textit{Buffer}_u$ means the set of data in epoch $u$'s \textit{first} buffer, and $\mathbf{card}$ means the cardinality of a set.
Note that $N_{u,v}$ cannot be the same as $N_{v,u}$ by definition.
Given a buffer size $|\textit{Buffer\hspace{1pt}}|$, the first buffer means the set of first $|\textit{Buffer\hspace{1pt}}|$ elements in epoch $u$, and so forth. To find an optimized solution to schedule the order of epochs, we have 
\begin{equation}
    \min \left(\text{cost}_\text{I/O} \right) = \min \Big( \sum^p N_{u,v} \Big),
\end{equation}
where $p$ is a path and $\text{cost}_\text{I/O}$ denotes the modeled total cost of path $p$.
Thus, the problem now is to find the shortest path in graph $G$ such that each vertex in the graph is visited exactly once.

This problem is similar to the well-known Traveling Salesman Problem (TSP)~\cite{flood1956traveling}. The differences are: (1) It does not need to go back to the origin vertex, which is defined as path-TSP and proved to be NP-complete~\cite{TSP_npc}; (2) The edges are bi-directional in terms of weights. 
Leveraging heuristic algorithms can enable a near-optimal solution to this problem.

\textbf{Implementation Details}
We utilize Particle Swarm Optimization (PSO) to solve the TSP problem.
PSO \cite{pso} is suitable for providing the near-optimal solution for this problem through observation, and it takes a graph as the input and outputs the shortest path that visits each vertex exactly once.

Specifically, PSO is an approximation algorithm that originated from the social behavior of swarms. 
First, each agent starts from a random vertex within the graph, and its fitness function is defined. 
The output of the fitness function refers to the distance between the current solution of the agent and the best solution among agents~\cite{shi2007particle}. 
Second, the swarm of agents moves synchronously so that the local-best solution is aggregated and updated to replace the global best solution at each iteration. 
Finally, the algorithm will terminate if all the vertices in the graph have been visited and will output the shortest path as well as the length of the shortest path.

\subsubsection{Boosting Data Locality}\hfill\\
\textbf{Limitations of Existing Work}
DeepIO~\cite{zhu2018entropy} tries to maximize the data reuse by increasing the data locality. It limits the 
sample shuffling within each device so that the buffered data can ultimately be reused. However, this approach 
seriously hurts the randomness and sharply reduces the training accuracy.
This limits DeepIO being applicable in some cases, because (1) For some tasks like image classification on Imagenet, the accuracy is over 90\%, thus trading off training accuracy is not a good choice. (2) The relationship between randomness and training accuracy is unknown for many neural networks and datasets, thus, we cannot know if it is worth utilizing the framework until trying it for each network and dataset.

\textbf{Observation}
To boost the data locality without sacrificing training accuracy, we design our optimization based on a key observation: with the determined shuffled
index list, changing the access order and node-to-sample mapping within a global batch will still result in the same gradient after synchronization (as proved in
~\cite{yang2019accelerating}).
Here, the node-to-sample mapping is defined as the relationship of which data sample to access in training on which node.
Since the dataset is fully randomized in each epoch, the node-to-sample mapping will vary across epochs. 
It incurs extra data loading from the file system or point-to-point communication to exchange the needed data samples.
Meanwhile, generating the mini-batch from locally buffered data samples significantly affects the training accuracy \cite{zhu2018entropy}.
Luckily, data parallelism is a synchronous training mechanism that averages the local gradients computed from mini-batches of data.
These mini-batches from multiple devices are together named the global mini-batch or global batch.
Specifically, changing the training order of data samples within a global batch does not affect the gradient results after synchronization \cite{yang2019accelerating}.
Consider the following update function in data parallelism:
\begin{equation}\label{eq:acc}
    w_{t+1} = w_t - \textstyle \eta{N_B}^{-1}\sum^N_{k=1} \sum_{x \in B_{k,t}} \nabla f(x,w_t)
\end{equation}
where $w_t$ denotes the weight at iteration $t$, $\eta$ denotes the learning rate, $N$ is the number of devices and $B_{k,t}$ is the mini-batch on device $k$ of iteration $t$, and $\nabla f(x,w_t)$ is the local gradient.
From the equation, we can see that so long as the data samples in a global batch ($\sum^N_{k=1} \sum_{x \in B_{k,t}} \nabla f(x,w_t)$) are trained together, changing the node to sample mapping within the global batch produces the same gradient result after gradient averaging.
Thus, we can leverage this characteristic of data parallelism to achieve a higher in-memory data reuse rate without affecting the training accuracy.

\textbf{Proposed Optimization}
With the key observations mentioned above, we are able to re-arrange the node-to-sample mapping to optimize data locality and thus improve the data reuse rate.
Specifically, we propose to improve the data reuse rate by updating the node-to-sample mapping based on locality, instead of requesting intra-node data movement via point-to-point communication, as shown in Figure~\ref{fig:overview} (c). Through offline computation, we can acquire the knowledge of node-to-sample mapping in the current epoch.
With this information, we can update the mapping relationship within each global mini-batch for the next epoch to benefit data reuse.
For example, for the original shuffle list, sample $s$ is trained on epoch $e_i$ on node $n_x$. For the next epoch $e_{i+1}$, it should be trained in step $j$ on node $n_y$. We arrange the access order to make the sample $s$ trained on node $n_x$ still on epoch $e_{i+1}$ and step $j$.


The node-to-sample mapping optimization in \TheWork{} is inspired by an interesting thought: How do you identify that the color you see is the same as others? 
For example, everyone names colors like ``Red,\!'' ``Green,\!'' and ``Blue,\!'' but how do people know that they see the same color in their vision? 
One answer can be: Learning name-to-color mapping is enough. 
It is similar to the data locality problem in distributed training in that we only need to remap the samples and nodes within a global batch.

\textbf{Implementation Details}
The data locality optimization is implemented by re-arranging the access order of each epoch according to its previous epoch. 
By doing so, we can avoid node-to-node communication.  
Specifically, we only need to arrange the access order within the global mini-batch by 
computing the samples accessed and stored in the last epoch, and remapping the buffered samples to the corresponding node for the coming epoch. 
The rest of the data samples will be evicted during runtime after the buffer is populated because their interval between the last accessed time and the next access time is large.
By doing so, we efficiently utilize the limited buffer storage space and achieve maximum data reuse.

\begin{figure}[t]
    \centering
    \includegraphics[width=0.8\linewidth]{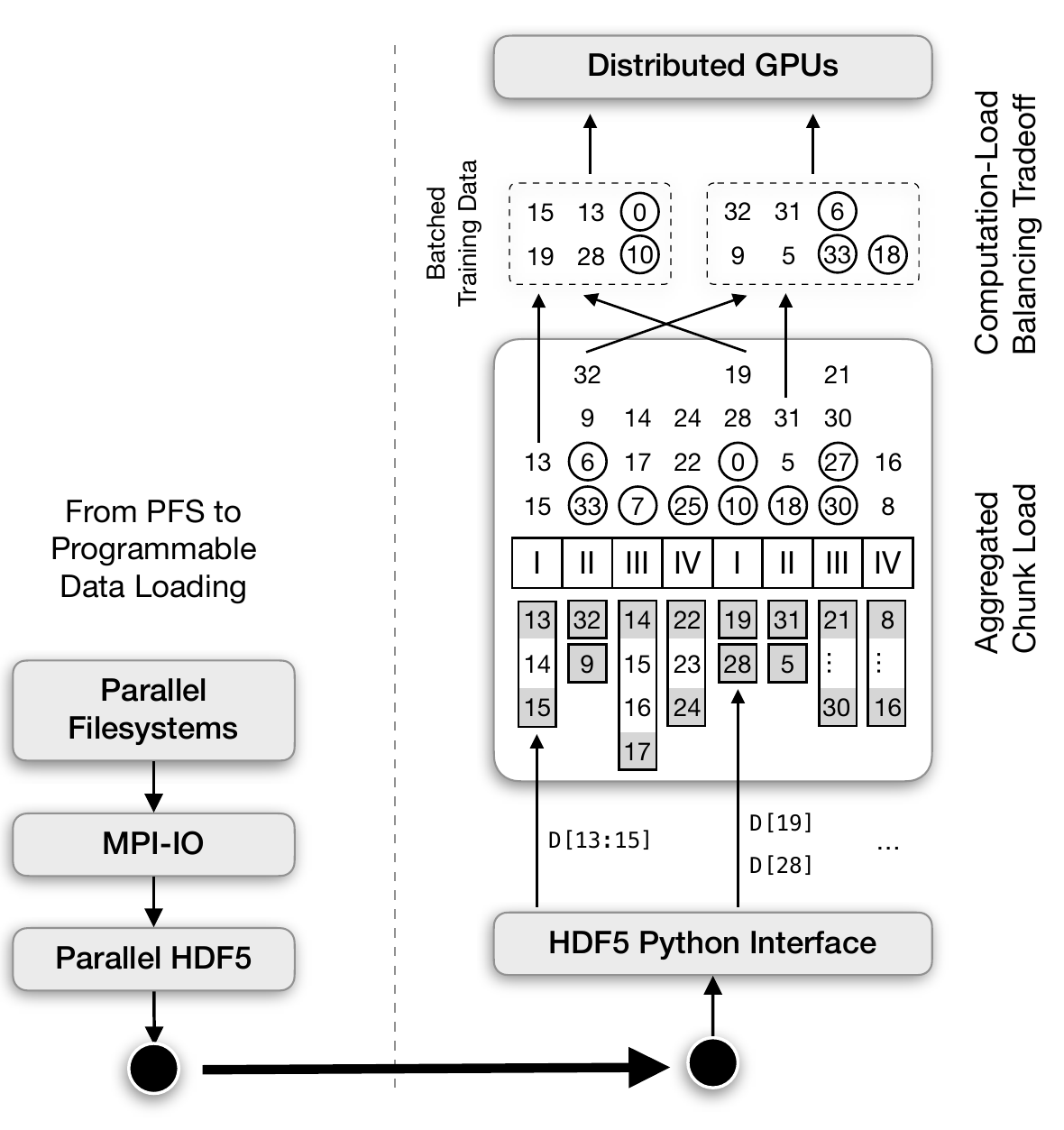}
    \caption{Data loading from PFS to GPUs using our proposed runtime buffering strategy with two optimizations. 
    }
    \label{fig:runtime}
\end{figure}

\subsection{Loading Workload Balancing}
\textbf{Limitations of Existing Work}
There are a few existing works that try to solve the second (imbalance) issue. Locality aware~\cite{yang2019accelerating} utilizes intra-node communication to balance the data loading during runtime.
By doing so, it has to schedule the sending source and receive target for each data sample which is an NP-Complete problem and can only be approximately solved.
This limitation will lead to performance loss since we still have extra I/O overhead
when the approximation algorithm is not accurate.
Lobster~\cite{liu2022lobster} tackles the imbalance problem by re-assigning preprocessing threads for data loading. It essentially trades the thread imbalance for a balanced loading workload.
In surrogate training, since data loading takes the majority of the training time, the thread 
imbalance will exist during most of the training process, and thus its system overhead will be amplified.
Besides, surrogates, like PtychoNN, do not require complex data preprocessing and data augmentation, 
hence cannot directly utilize Lobster.

\begin{figure}[b]
    \centering
    \includegraphics[width=\linewidth]{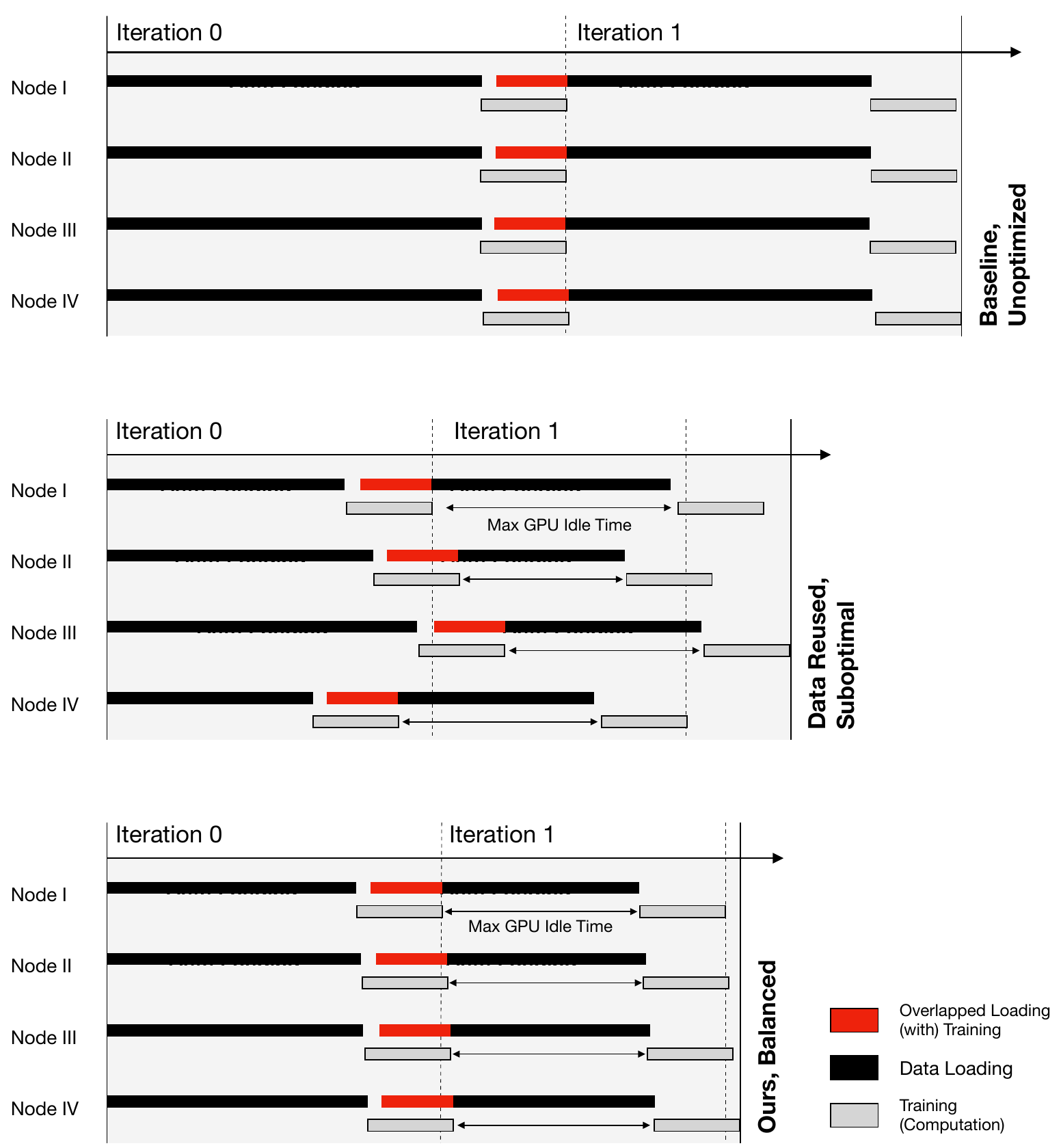}
    \caption{An example illustrating performance improvements with data load balancing.}
    \label{fig:bal_imp}
\end{figure}

\textbf{Observation}
In order to overcome the imbalance issue, we leverage an observation that is important for our load-balancing optimization:
computational imbalance is less harmful to overall training time compared to the load imbalance.
In data parallelism, due to the random shuffle, the number of data samples that can be reused on each node is not the same.
In other words, the number of samples that need to be loaded from the PFS is different across nodes, resulting in imbalanced I/O.
Furthermore, this imbalance causes a large amount of idle time for GPUs as they must wait for the data to be loaded from the file system to the host memory, formatted into a batch, and transferred to GPUs. 
The importance of data load balancing to training performance is illustrated in Figure~\ref{fig:bal_imp}. We can see that when the data loading time is significantly larger than the computation time, prefetch cannot overlap with the computation. 
After the first optimization in the middle example, the number of samples loaded from the parallel file system is reduced, thus the I/O time is reduced. 
Meanwhile, the data imbalance is introduced, which limits the data loading efficiency. 
In the bottom example, after data load balancing, the overall training time is reduced.
\begin{figure}[t]%
    \centering
    \includegraphics[width=\linewidth]{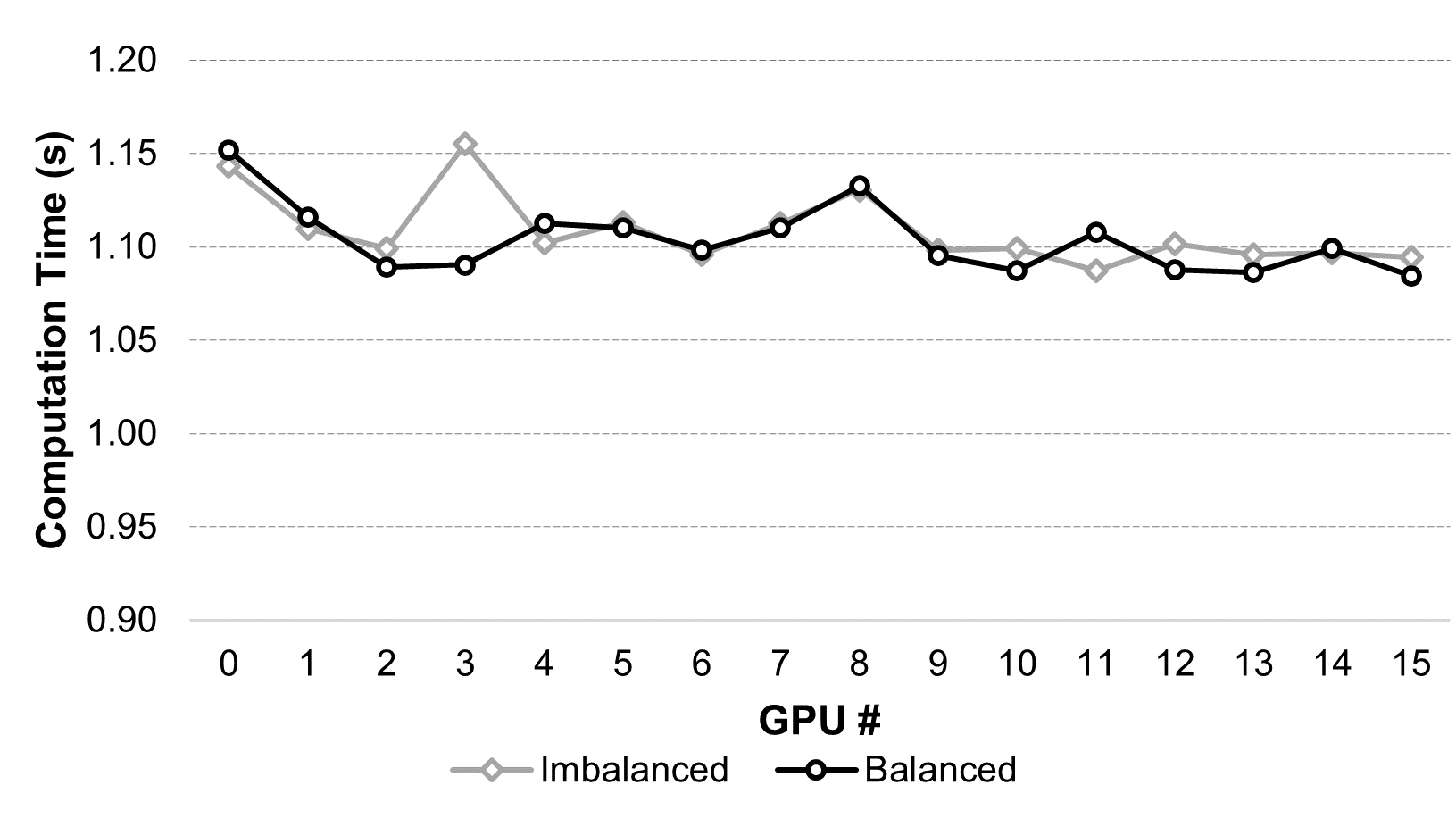}
    \caption{Comparison of computation time on each GPU between original training and our solution with the load-balancing trade-off, demonstrating that the imbalanced batch size does not affect performance. 
    }%
    \label{fig:comp_balance}%
\end{figure}

The voyage to explore possible data load balancing optimization designs is stormy.
Fortunately, when we performed micro-benchmarking experiments to observe the computation time for balanced and imbalanced training batch sizes, we discovered that the computation time does not differ much between the two cases.
Specifically, 16 A100 GPUs are used in the experiments, and we first fixed the batch size for each GPU and trained on that setting;
Then we trained with modified batch sizes where each batch size is the defined batch size minus the rank index of each GPU.
We observe that varying the batch size when training results in a similar computation time compared to the fixed batch size training, as shown in Figure~\ref{fig:comp_balance}.
The black line shows the average computation time of 10 epochs when the batch size is set to 64. 
The gray line shows the average computation time of 10 epochs when batch size is set to $|\textit{batch}| = 64-\textit{rank}$. 
Note that the computation time of both balanced and imbalanced solutions are relatively close to each other, considering some system variance between GPUs.
This means we can trade off the compute balance for the load balance to improve the overall performance.

\textbf{Proposed Optimization}
With the key observation mentioned above, we design the data load balancing optimization that trades computational balance for load balance.
Specifically, to balance the data loading across nodes, our goal is to let each node load a similar amount of data during every step.
After designing a buffer for data reuse, the data samples within a mini-batch have two states: (1) to fetch from the PFS, and (2) to fetch from the in-memory buffer.
The number of data samples in the state (1) on multiple GPUs is not the same, thus the data loading is not balanced for the data parallelism.
From our observation, we design an optimization to distribute the number of data samples evenly in the state (1) across multiple GPUs.
By doing so, the training batch sizes on the GPUs are different.
As shown in equation \ref{eq:acc}, since we did not change the global batch, this optimization will not affect training accuracy.
Furthermore, we observe that changing the training batch size does not affect computation time much, and we demonstrate the importance of data load balancing in Figure~\ref{fig:bal_imp}.
This is the trade-off: we diminish the computation balance to achieve load balance.

\textbf{Implementation Details} 
During offline scheduling, we compute the indices of the data that need to be fetched from the PFS.
During runtime, we use those indices to perform the computation balance trade-off, as shown in Figure~\ref{fig:runtime}.
Once the mini-batch of data is ready on a GPU, the computation part of the training begins.
\subsection{Aggregated Chunk Loading}
\label{sec:aggregated}

HDF5~\cite{hdf5} is an I/O library designed for HPC applications and achieves high read/write throughput.
However, we observe that the throughput of data loading with parallel HDF5 is low in our training. Specifically,
while loading the same amount of data in total, random access to many small files incurs higher loading time compared to chunked loading. This raises the question of how different access patterns can affect the data-loading performance. Specifically, there are four flavors of access patterns, 
(1) \textit{random access}, in which a process reads one sample randomly until all samples in the dataset have been accessed once,
(2) \textit{sequential-stride access}, in which a process iteratively reads samples with a fixed stride,
(3) \textit{chunk-cycle loading}, in which a process loads samples one by one in its assigned chunk,
and (4) \textit{full-chunk loading}, in which a process loads multiple consecutive samples in its assigned chunk in one go.
\begin{figure}
    \centering
    \includegraphics[width=\linewidth]{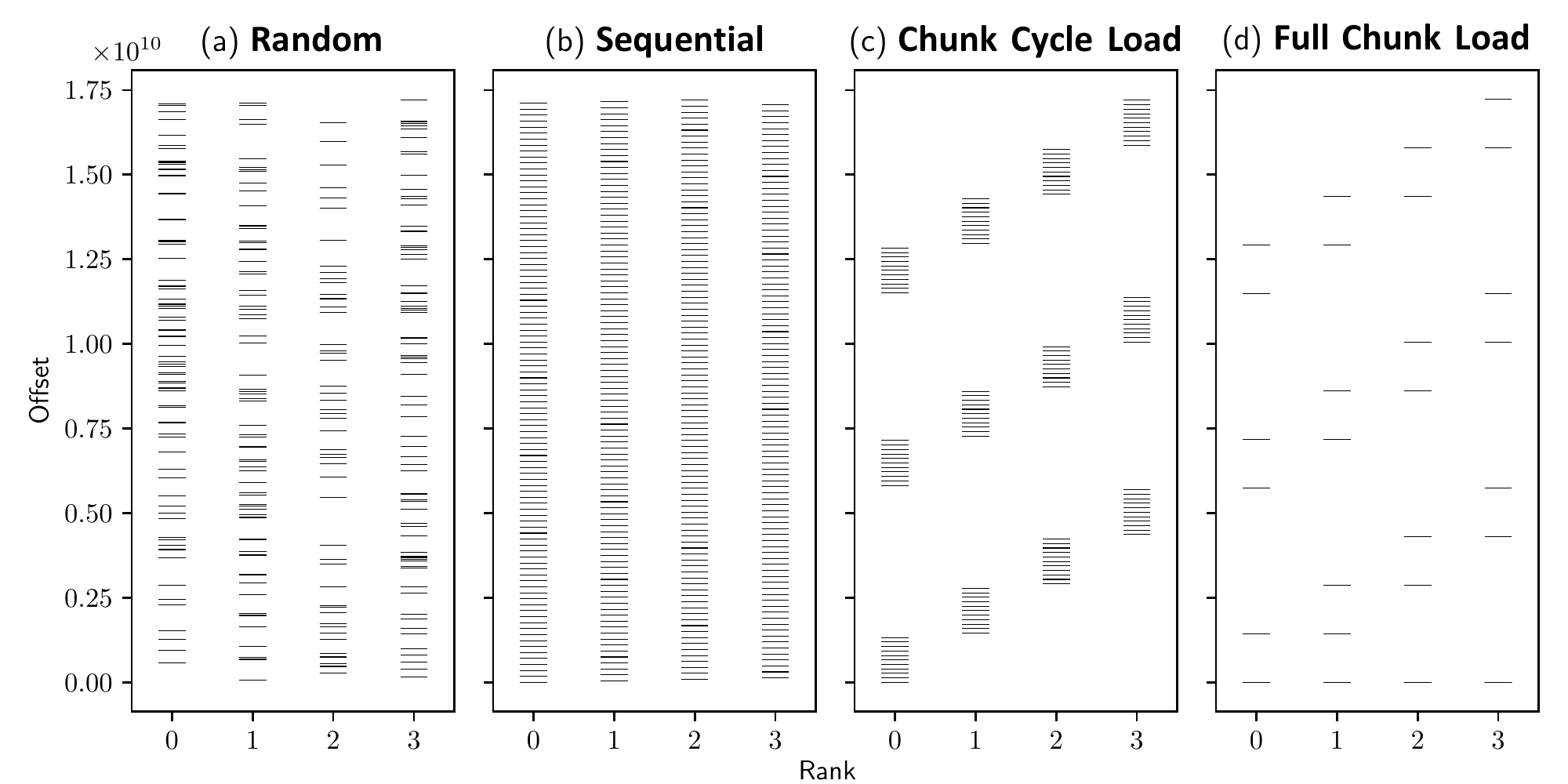}
    \caption{An example of offsets visited in an HDF5 file with different access patterns.}
    \label{fig:acc_patt}
\end{figure}

\begin{table}[b]
    \TableStyle
    \caption{I/O time (in seconds) of different access patterns.}
    \begin{tabular}{@{} >{\bfseries}l rrr  @{}}
        \toprule
        \bfseries Pattern        & \bfseries Time & \bfseries Norm'ed$\phantom{\times}$ & \bfseries Speedup \\
        \midrule
        {Random Access}            & 645.864        & 203.42$\times$                      & 1.00$\times$      \\ 
        {Sequential Stride Access} & 84.421         & 26.59$\times$                       & 7.65$\times$      \\ 
        {Chunk Cycle Loading}      & 30.537         & 9.62$\times$                        & 21.15$\times$     \\ 
        {Full Chunk Loading}       & 3.175          & 1.00$\times$                        & 203.42$\times$    \\ 
        \bottomrule
    \end{tabular}
    \label{table:I/O_time}
\end{table}

We set up a microbenchmark to profile the I/O performance under different access patterns. 
Specifically, we utilize the record profiling tool~\cite{recorder} to visualize the I/O activity with the accessed offsets in the HDF5 file.

\textbf{Observation} Figure~\ref{fig:acc_patt} shows the trace of different patterns.
The random access pattern results in uneven distribution of offsets on a single process. The offset reveals the same sequential feature for sequential access and chunked cycle loading. 
Table~\ref{table:I/O_time} shows the measured I/O time; 
we can see that sequential access is faster than random access since each process spends a lot of time seeking the data for random access. 
On the other hand, chunk cycle loading takes less time than sequential access because consecutive access achieves better I/O performance than stride access in seeking the following data. 
Overall, full chunk loading achieves the best performance with a 203$\times$ speedup compared with random access.

\textbf{Proposed Optimization}
Our optimization strategy is twofold: (1) we rearrange the data loading to be as consecutive as possible, and (2) changing the access order within the mini-batch maintains the randomness, equivalent to the original random shuffle. We first find the optimal chunk size from the micro-benchmark experiment. Then we determine if the current and the next sample to load can be combined in the same chunked loading, based on the known indices from our proposed offline scheduling.

\textbf{Implementation Details}
PyTorch provides the \verb|__getitem__| function in the custom dataset option that accesses one data sample at a time~\cite{paszke2019PyTorch}, then the collate function is called to collate the data samples into a batch of data. We utilize the interface by (1) sorting the indices of the samples to be loaded in a mini-batch, (2) determining which samples can be loaded in a chunk, and (3) loading multiple samples in chunks, as shown in Figure~\ref{fig:runtime}.
\section{Performance Evaluation}
\label{sec:evaluation}
In this section, we answer three main questions:
(1) What is the performance of \TheWork{} compared to state-of-the-art approaches on different datasets with different system setups? (\S\ref{sec:io_improvement})
(2) What is the contribution of each optimization to \TheWork{}? (\S\ref{sec:ablation})
(3) What is the end-to-end training performance with \TheWork{}? (\S\ref{sec:end2end})
\subsection{Experimental Setup}\label{sec:datasets}




Our framework is implemented using PyTorch.
In our experiments, we use DGX-A100 GPU nodes with Lustre parallel file system.

\begin{figure*}
    \centering
    \includegraphics[width=\linewidth]{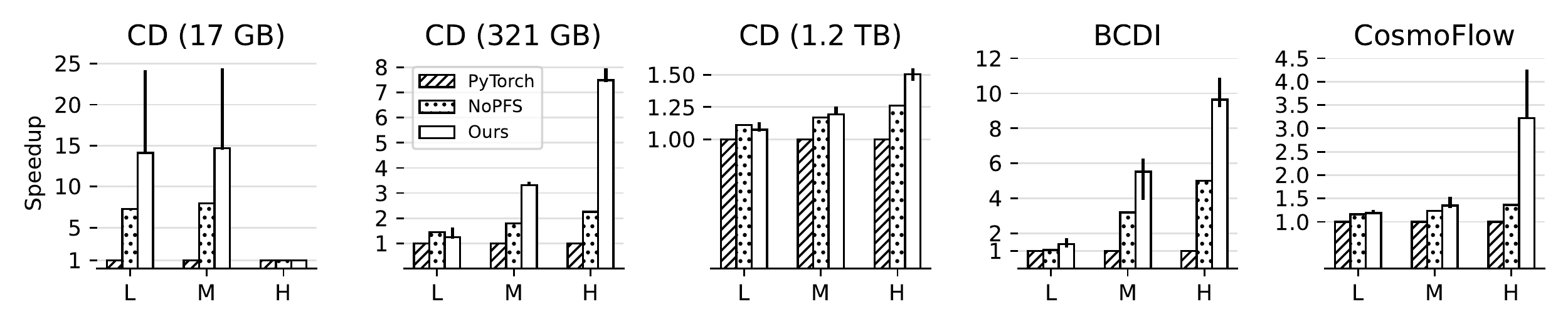}
    \vspace{-6mm}
    \caption{Comparison of data-loading performance on three systems using five datasets.}
    \vspace{-2mm}
    \label{fig:dl_impv}
\end{figure*}

\textbf{Baselines}
We use two baselines for comparison, namely PyTorch DataLoader and NoPFS.
PyTorch Dataloader~\cite{ddp} is the data loading framework that is widely used by the industry and academia.
It is designed to load the data for training and validating the neural networks implemented by PyTorch.
NoPFS~\cite{dryden2021clairvoyant} is an existing data-loading framework that claims to outperform state-of-the-art approaches, including PyTorch DataLoader, DeepIO~\cite{zhu2018entropy}, LBANN~\cite{jacobs2019parallelizing}, and Locality-aware~\cite{yang2019accelerating}.

\textbf{Datasets}
We evaluate \TheWork{} on five datasets. (1) Coherent Diffraction Dataset (CD) has 262,896 images (17 GB) in total, and the size of each image is 65KB. We also synthesize the dataset of two sizes (321 GB, 1.2 TB) to study the performance of using larger-scale datasets, which have 1,752,660 and 18,928,620 samples, respectively. (2) 3D X-ray Bragg coherent diffraction imaging (BCDI) Dataset has 54,030 samples in total, each sample is 3.1MB. (3) 3D Universe Simulation (CosmoFlow) Dataset has 63,808 samples (1 TB), each one being the 3D universe simulation data of size 17MB.

\textbf{Buffer sizes}
The performance of \TheWork{} is affected by a few factors, among which the buffer size $|\textit{Buffer\hspace{1pt}}|$ is the major one.
Thus, we need to study the improvement of \TheWork{} compared to the baselines using different buffer sizes.
Specifically, there are \textit{three} scenarios:
(1) dataset size $\leq$ local buffer size,
(2) local buffer size $<$ dataset size $\le$ total buffer size, 
(3) dataset size $>$ total buffer size.

\begin{table}[ht!]
    \TableStyle
    \caption{Number of GPUs used for different datasets. Buffer size is the number of GPUs multiplied by 8/16/40 GBs for low-/medium-/high-end systems, respectively.}
    \begin{tabular}{@{} >{\bfseries}l rrrrr  @{}}
        \toprule
        & \multicolumn{3}{c}{CD} &                 &                                                   \\
        \cmidrule(lr){2-4}
                    System     & 17 GB                  & 321 GB              & 1.2 TB              &      BCDI          &                   CosmoFlow  \\
        \midrule
        {Low-end}        & 2                      & 16                  & 32                  & 8              & 16                  \\
        {Midium-end}     & 2                      & 16                  & 32                  & 8              & 16                  \\
        {High-end}       & 2                      & 8                   & 16                  & 8              & 16                  \\
        \bottomrule
    \end{tabular}
    \label{tab:setup}
\end{table}

\subsection{Evaluation of I/O Improvement}\label{sec:io_improvement}



We first show the I/O improvement of \TheWork{} compared to PyTorch DataLoader and NoPFS on five datasets under three different system setups. 
Specifically, as shown in Table~\ref{tab:setup}, the low-/medium-/high-end systems have 8/16/40 GB buffer per GPU, respectively.
To simplify the problem, we assume one GPU per node to demonstrate the performance of \TheWork{} on distributed buffers.

The performance improvement of \TheWork{} compared to PyTorch DataLoader and NoPFS is shown in Figure~\ref{fig:dl_impv}.
Specifically, the bar shows the speedup compared to PyTorch DataLoader.
To compare \TheWork{} with PyTorch DataLoader, we measure data loading time across multiple runs and take the average, then we compute the speedup.
For NoPFS, we run their simulator that is implemented based on NoPFS's performance model and outputs the simulated data loading time of PyTorch DataLoader and NoPFS.\!\footnote{NoPFS only provides a simulator in its open-source repository.}
The general trend shows that as the aggregated buffer size increases, \TheWork{} achieves better speedup.
For example, on the five datasets, we observe that the speedup of \TheWork{} compared to PyTorch DataLoader is higher on high-end systems than on medium-end and low-end systems.
This is because high-end systems have a larger buffer size, which is consistent with our design.

Next, we present the evaluation results in three scenarios.

\textbf{(1) dataset size $\leq$ local buffer size:}
In this scenario, the entire dataset is loaded on each GPU.
For example, on the CD 17 GB dataset with the high-end system, NoPFS and \TheWork{} achieve no speedup because during the first epoch, all data samples are loaded from the PFS and for the following epochs, all data samples are loaded from the local buffer.

\textbf{(2) local buffer size $<$ dataset size $\leq$ total buffer size:}
In this scenario, \TheWork{} loads almost all the data samples from the local buffer, while NoPFS loads half of the samples from the local buffer and half from the remote buffers (on other nodes).
Therefore, \TheWork{} benefits from data locality optimization and aggregated chunk loading, so that \TheWork{} outperforms NoPFS.
For example, on the CD 17 GB dataset with the medium-end system, \TheWork{} achieves 14.1$\times$ average (up to 24.4$\times$) speedup compared to PyTorch DataLoader and 1.9$\times$ average (up to 3.3$\times$) speedup over NoPFS.
Also, on the BCDI dataset with high-end systems, \TheWork{} achieves 9.6$\times$ average (up to 10.9$\times$) speedup compared to PyTorch DataLoader and 1.9$\times$ average (up to 2.2$\times$) speedup over NoPFS.

\textbf{(3) dataset size $>$ total buffer size:}
In this scenario, \TheWork{} loads the data samples from the local buffer and the PFS, while NoPFS loads from the local buffer, the remote buffer, and the PFS.
Note that the PyTorch DataLoader does not use the buffer. 
Here, \TheWork{} outperforms NoPFS because it achieves a better data reuse rate and avoids inter-node data movement compared to NoPFS.
For instance, on the CD 321 GB dataset, CD 1.2 TB dataset, and CosmoFlow dataset, \TheWork{} achieves up to 7.96$\times$/1.55$\times$/4.25$\times$ speedup compared to PyTorch DataLoader, and up to 3.52$\times$/1.23$\times$/3.13$\times$ speedup over NoPFS, respectively.
Meanwhile, we observe the performance improvement limitation of two datasets (CD 321 GB and CD 1.2 TB) on the low-end system.
On the one hand, although the average speedup of \TheWork{} does not achieve performance improvement as high as NoPFS, our best run outperforms NoPFS in both cases.
On the other hand, the result shows that when the buffer size is relatively small compared to the dataset size, the performance improvement of \TheWork{} can be limited compared to NoPFS.
This is because when the buffer size is small and the per-sample size is small, the access order optimization cannot significantly improve the data-loading performance.
This shows the limitations of \TheWork{} and NoPFS, in that when the dataset is too large for the buffer, the improvement is not significant compared to PyTorch DataLoader.
However, the worst case for \TheWork{} is that the performance is close to NoPFS, since NoPFS does not re-arrange the access order and \TheWork{} does not incur extra overhead (inter-node communication).

In addition, a recent work, called Lobster~\cite{liu2022lobster}, develops a solution to address I/O load imbalances among GPUs during DNN training. However, we exclude Lobster from our comparison for three main reasons: (1) while our work focuses on data-loading performance, Lobster also optimizes pre-processing (such as decoding, shuffling, batching, and augmentation); (2) Lobster’s code is not available for testing (we discuss more details of Lobster in \S\ref{sec:related}); and (3) Lobster’s performance improvement over NoPFS is up to 1.2$\times$, while \TheWork{} can improve the data-loading performance by 2.37$\times$ over NoPFS.

\subsection{Evaluation of Each Optimization}\label{sec:ablation}
Next, we break down the performance improvement to demonstrate the contribution of each optimization.
Specifically, Optim\_1 is the access order optimization, Optim\_2 denotes data load balancing, and Optim\_3 is aggregated chunk loading.
We also implemented a buffer with the least recently used (LRU) strategy for PyTorch DataLoader (SOLAR--Optim\_1--Optim\_2--Optim\_3).

\begin{figure}[ht]
    \centering
    \includegraphics[width=\linewidth]{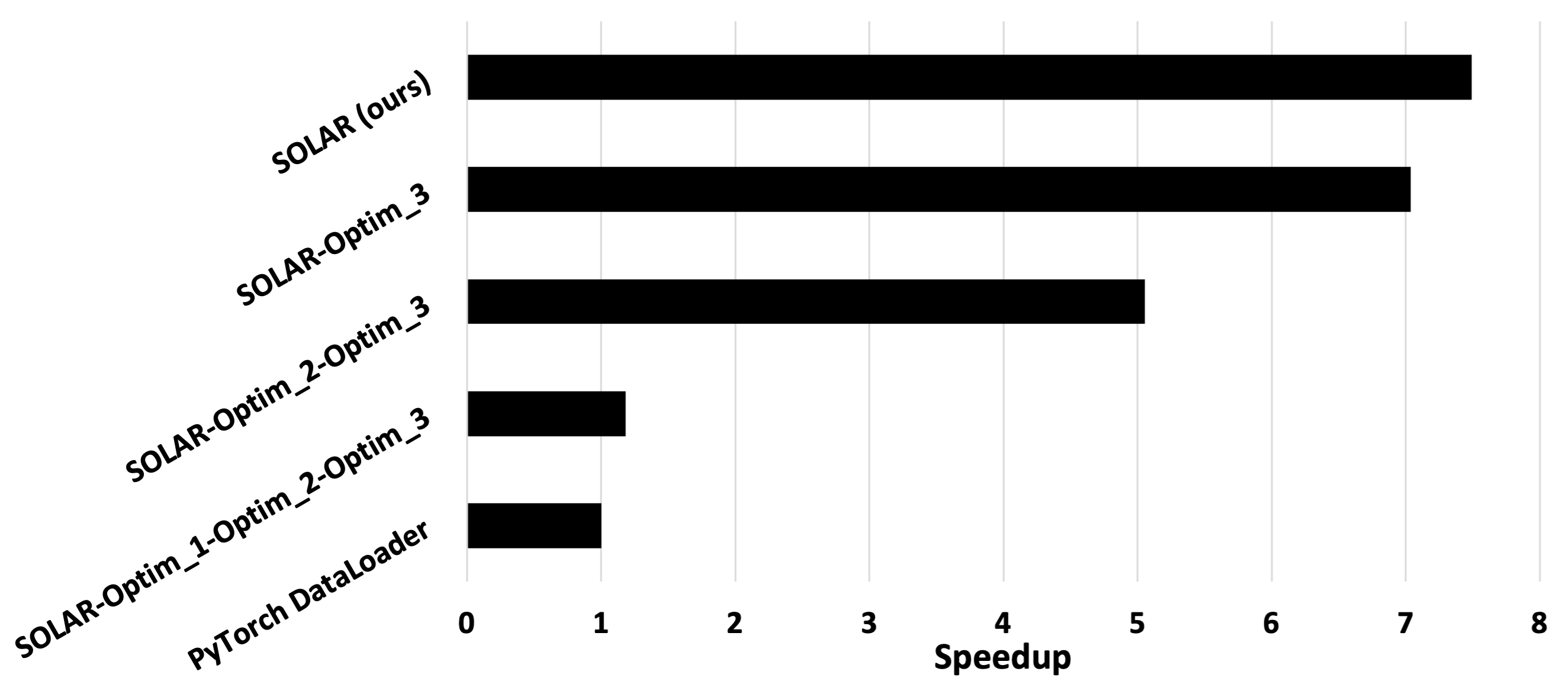}
    \caption{Performance breakdown for each optimization. Optim\_1 is access order optimization, Optim\_2 is data load balancing, and Optim\_3 is aggregated chunk loading.}
    \label{fig:breakdn}
\end{figure}

Figure~\ref{fig:breakdn} shows the performance breakdown for each optimization.
First, we observe that compared to PyTorch DataLoader, PyTorch DataLoader + LRU buffer has a 1.2 $\times$ speedup.
That shows the effect of using a buffer for data reuse.
Second, we can see that the access order optimization has a significant effect on performance improvement because by scheduling the access order we are able to maximize the data reuse;
furthermore, compared to PyTorch DataLoader + LRU buffer, the access order optimization achieves the most improvement.
This is because by adjusting the access order, \TheWork{} is able to ``intelligently'' improve the data reuse rate more when using a larger buffer size while PyTorch DataLoader + LRU buffer does not efficiently utilize the in-memory buffer space for data reuse.
Third, data load balancing optimization further reduces the data loading time because the parallel reading idle time is reduced when we balance the data loading workload.
Fourth, the aggregated chunk loading improves the performance further, up to a cumulative 7.5$\times$ speedup, illustrating that the aggregated chunk loading is effective.
In this experiment, the chunk size is set to 15.\!\footnote{
The chunk size $|\textit{chunk\hspace{.5pt}}|$ is a threshold from benchmarking.
E.g., $|\textit{chunk\hspace{.5pt}}|=15$ means loading samples $i$ to $i+14$ in a single chunk is faster than loading them separately.}

\begin{figure}[ht!]
    \centering
    \vspace{-2mm}
    \includegraphics[width=.9\linewidth]{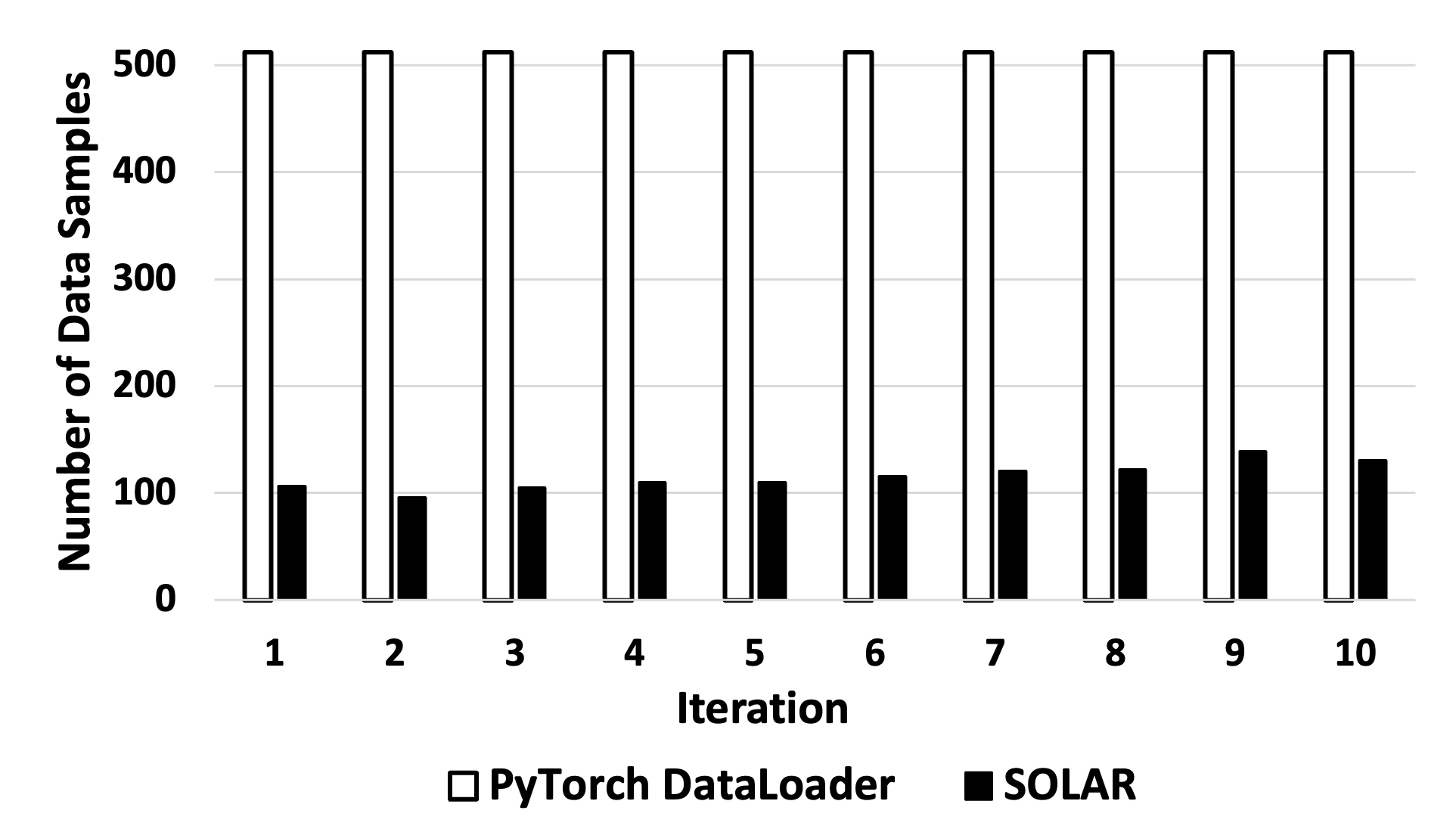}
    \caption{Comparison of the number of samples loaded from the PFS between PyTorch DataLoader and \TheWork{}.}
    \vspace{-2mm}
    \label{fig:aoo}
\end{figure}

\textbf{Access Order Optimization}
First, we show the number of data samples loaded from the PFS (``numPFS'') to demonstrate the contribution of access order optimization (``Optim\_1''), as shown in Figure~\ref{fig:aoo}.
The batch size is set to 512, and we take the maximum numPFS value from 16 GPUs.
Note that for PyTorch DataLoader, there is no reuse policy and thus it loads 512 samples per GPU on each iteration.
From the figure, we can see that numPFS is reduced up to 4.9$\times$ with the access order optimization.
Figure~\ref{fig:breakdn} shows the performance improvement from the reduction of numPFS.

\begin{figure}[t]
    \centering
    \includegraphics[width=\linewidth]{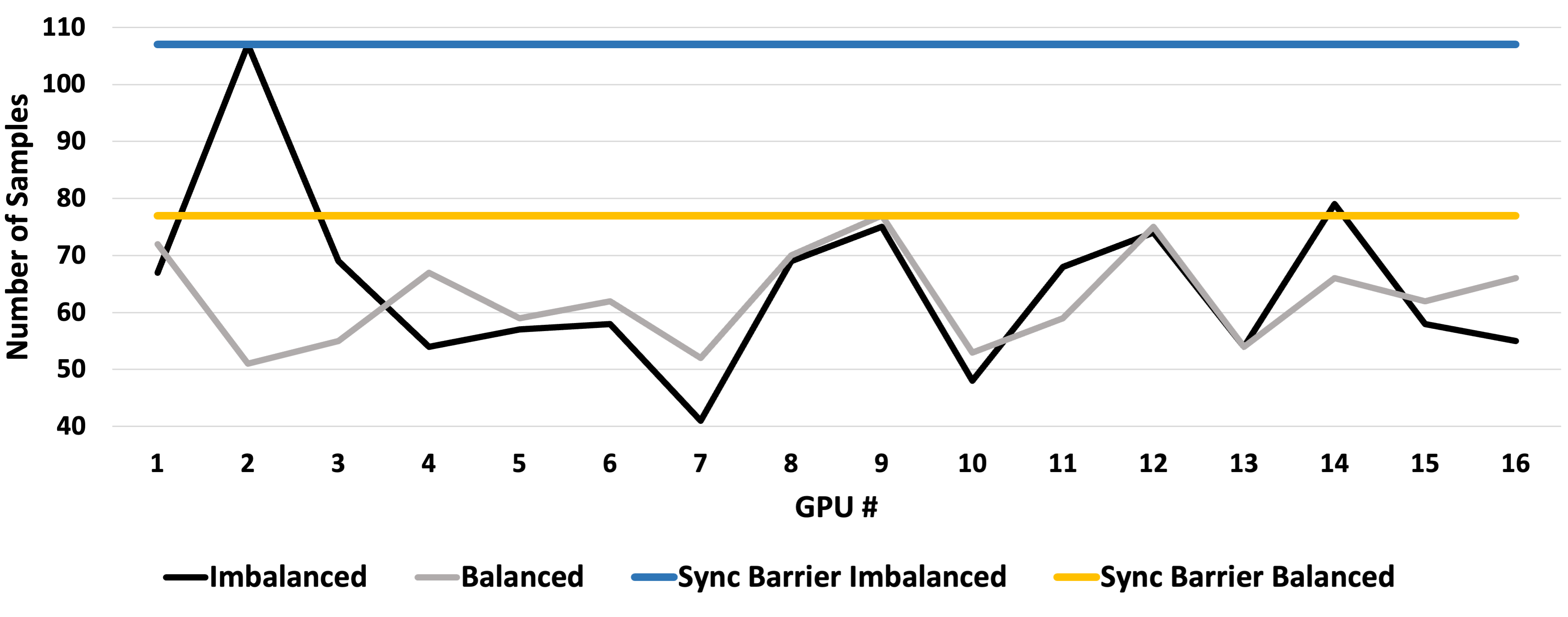}
    \caption{Comparison of the number of samples loaded from the PFS before and after load balancing optimization.}
    \label{fig:ld_bal}
\end{figure}

\textbf{Data Load Balancing}
Next, we demonstrate the contribution of load balancing.
We measure numPFS on each GPU (16 GPUs in total), as shown in Figure~\ref{fig:ld_bal}.
Specifically, the load imbalance is caused by the different number of reused data samples on each GPU.
In the imbalanced case, GPU 7 needs to load just 41 samples from the PFS, while GPU 2 needs to load 107.
All other GPUs need to wait for GPU 2 to finish loading (i.e., sync barrier imbalanced in Figure~\ref{fig:ld_bal}), thus causing idle time as a consequence.
In the balanced case, the GPU idle time is reduced.
The finish line (i.e., sync barrier balanced) is shown in Figure~\ref{fig:ld_bal}.
Thus, after load balancing the data loading is improved by 1.39$\times$ compared to the imbalanced case, and it is consistent with the speedup of I/O time shown in Figure~\ref{fig:breakdn}.

\begin{figure}[ht!]
    \centering
    \vspace{-4mm}
    \includegraphics[width=0.8\linewidth]{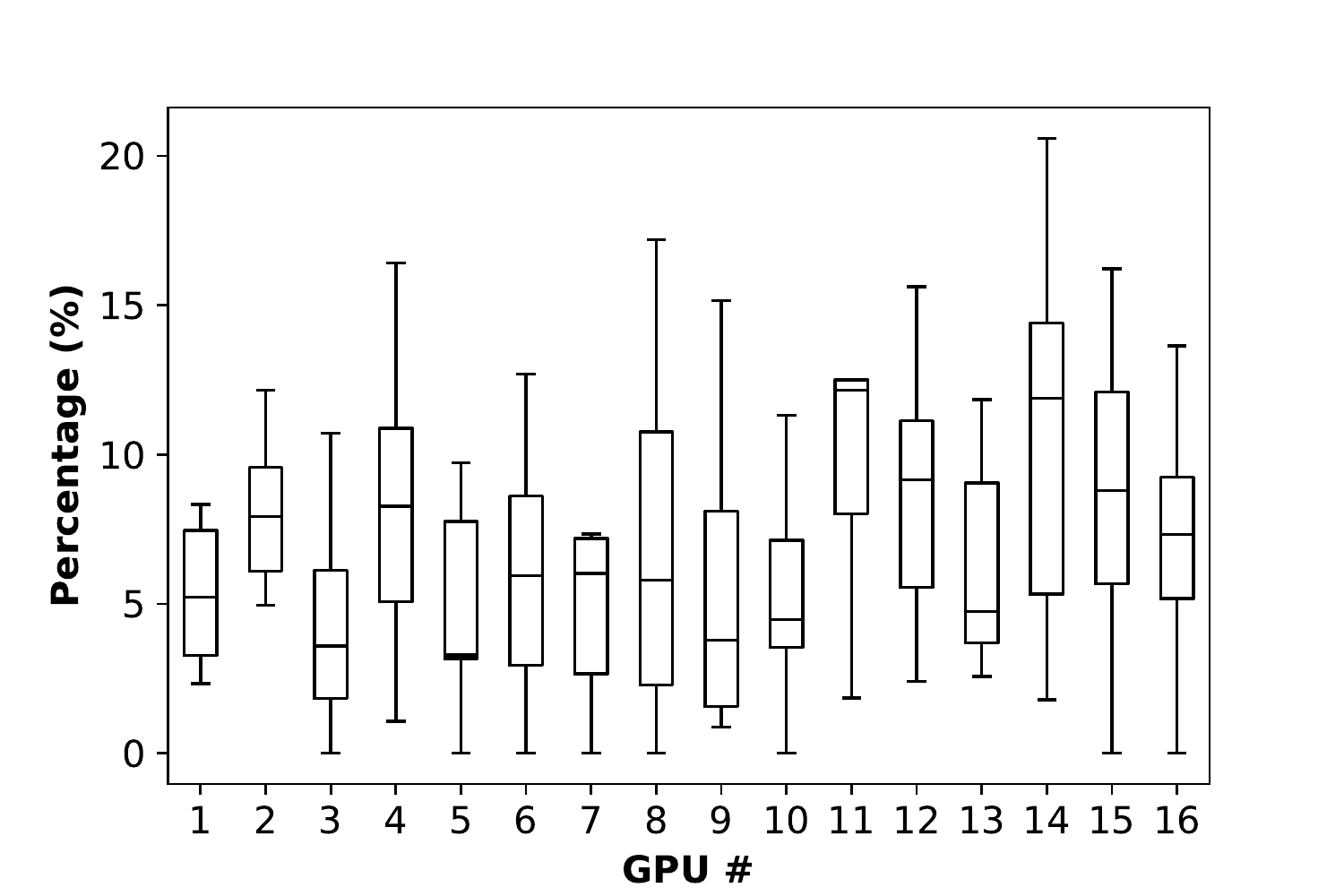}
    \caption{Percentage of data samples that can be loaded in chunks in different training runs.}
    \vspace{-4mm}
    \label{fig:ck_pct}
\end{figure}

\textbf{Aggregated Chunk Loading}
Lastly, we demonstrate the contribution of aggregated chunk loading.
The number of data samples that can be loaded in chunks is measured across multiple runs on each GPU, then divided by numPFS to compute the percentage. 
As shown in Figure~\ref{fig:ck_pct}, about 7\% of data samples on average (up to 20.6\% in the best case) can be loaded in chunks.
Meanwhile, the worst case is that no data samples can be loaded in chunks.
Note that in the worst case, this optimization does not harm the performance because there's no extra overhead.

Furthermore, we see the potential of this optimization to be applied to distributed training tasks that do not require full randomization.
For example, MLPerfHPC performs intra-node shuffle for DeepCAM~\cite{farrell2021mlperf}.

\subsection{Evaluation of End-to-End Training}\label{sec:end2end}
\begin{figure}[ht!]
    \centering
    \includegraphics[width=\linewidth]{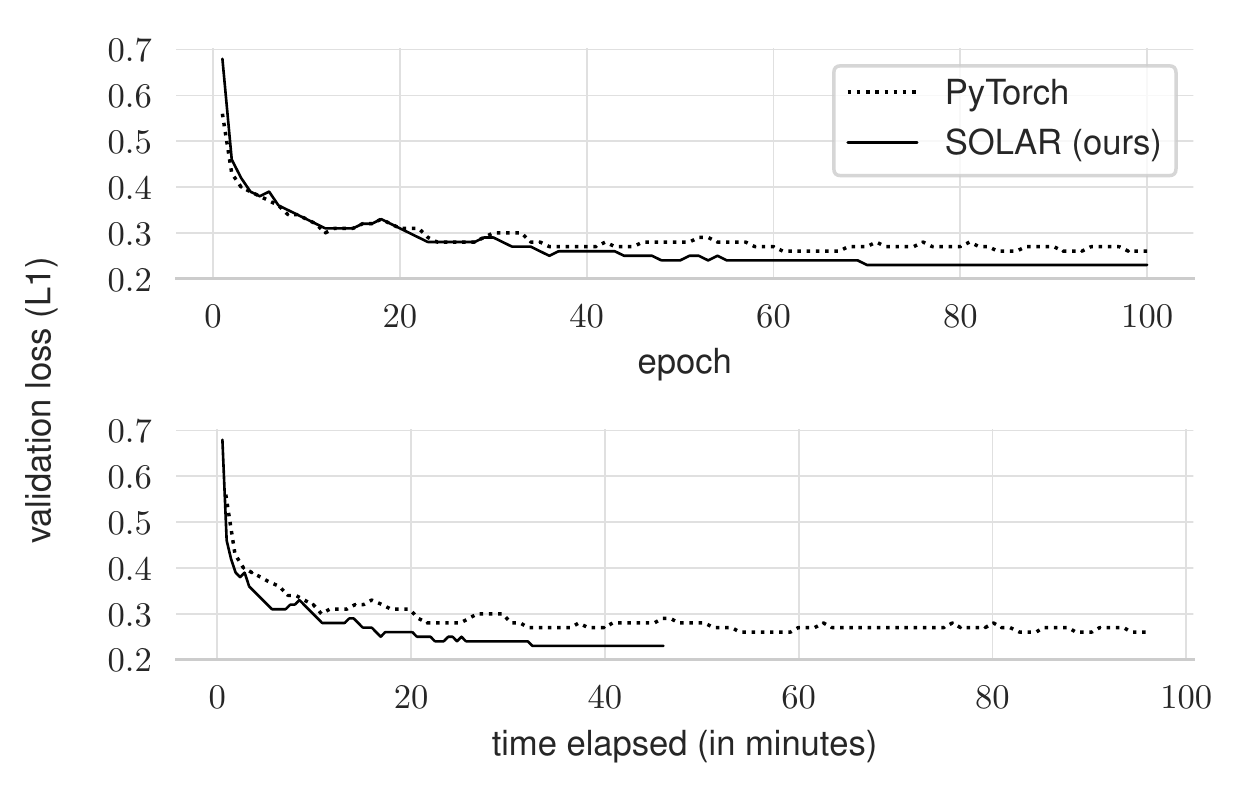}
    \caption{Training accuracy vs. time with PyTorch DataLoader method and our \TheWork{} on PtychoNN.}
   \label{fig:acc_comp}
\end{figure}

\begin{figure}[b]
    \centering
    \includegraphics[width=\linewidth]{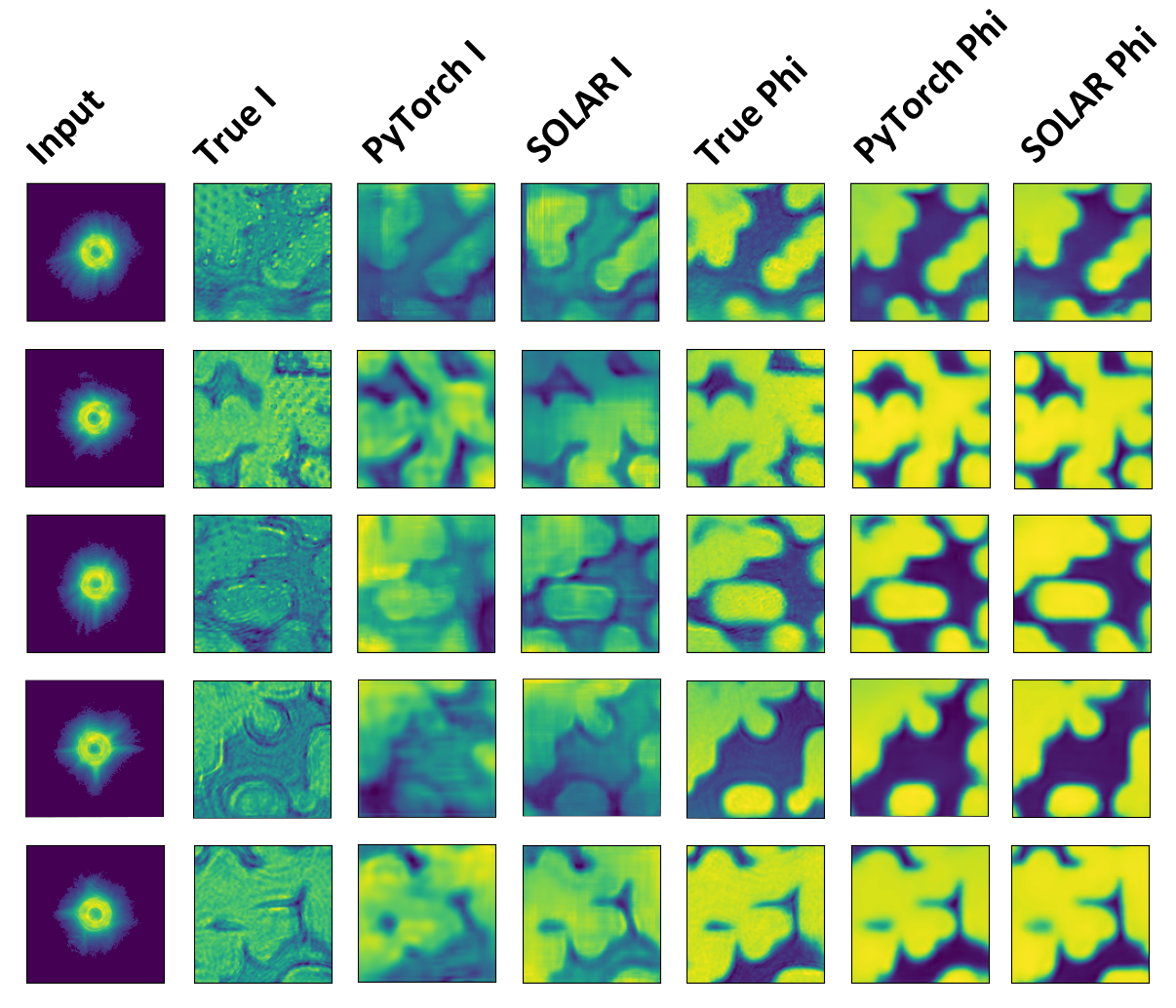}
    \caption{Comparison of outputs generated by PtychoNNs trained with PyTorch DataLoader (PyTorch) and \TheWork{}. ``True'' is the ground truth targeted in training. ``I'' and ``Phi'' denote amplitude and phase, respectively.}
    \label{fig:acc_refer}
\end{figure}
In this section, we evaluate the end-to-end training performance of \TheWork{} on the (CD 321GB) dataset using the high-end system.
Also, we show that the epoch order optimization can affect the training accuracy on PtychoNN.
We also performed an ablation study to demonstrate the options of \TheWork{}, which will be discussed in \S\ref{sec:discussion}
Specifically, we chose the 17 GB dataset to perform the accuracy study because the 321 GB and 1.2 TB datasets are synthesized from the 17 GB dataset by repeating the same images.
In order to reflect the realistic case where the dataset cannot fit in a single device, we set the local buffer size to 4.25 GB.
Figure \ref{fig:acc_comp} shows the end-to-end training comparison between the PyTorch DataLoader and \TheWork{}.
We can see that \TheWork{} not only improves the training performance but also achieves lower validation loss as well.
The time-to-solution speedup is 3.03$\times$ compared to PyTorch DataLoader.

After the model is trained, we perform the reference task using the test data.
Figure \ref{fig:acc_refer} shows the reference result compared to the ground truth labels (True I and True Phi), where we can see that \TheWork{} does not degrade the quality of image reconstruction and it can generate clear shapes of the amplitude (SOLAR I) and phase (SOLAR Phi).
We also surprisingly find that compared to the PyTorch DataLoader amplitude (PyTorch I) and phase (PyTorch Phi), \TheWork{} even improves the image quality in some cases.

\subsection{Detailed Analysis}\label{sec:discussion}
In this section, we answer two questions about \TheWork{}: (1) If changing the epoch order affects the randomness, what will the performance be if the epoch order optimization is not used? (2) How imbalanced will the \textit{training batch size} be after trading computation balance for load balance?

\textbf{Effect of epoch order optimization}
We perform a study to demonstrate the improvement of \TheWork{} with and without the Epoch Order Optimization (EOO).
As demonstrated in \S\ref{sec:design}, the data locality optimization, loading workload balancing, and aggregated chunk loading do not affect the training accuracy because they do not change the training results after synchronization.
However, when changing the epoch order there is a concern that it changes the randomness.
On the one hand, the term ``randomness'' is not quantized and is challenging to measure.
On the other hand, the relationship between randomness and training accuracy is unknown to the deep learning community.
To this end, we perform the study here to demonstrate the effect of re-arranging epoch order on data-loading performance.
First, to prove from the experiment that rearranging the epoch order improves the data-loading performance, we implemented a buffer that uses the Least Recently Used (LRU) buffering strategy for PyTorch DataLoader.
We observe that applying the EOO on PyTorch DataLoader + LRU improves the performance by 25.6\% compared to that without the EOO.
Second, we perform the experiment to study the effect of EOO on \TheWork{}.
We observe that when using the EOO in \TheWork{}, the performance improved by 59.4\% compared to not using the EOO.
\begin{figure}[]
    \TableStyle\scriptsize
    \centering
    \includegraphics[width=0.9\linewidth]{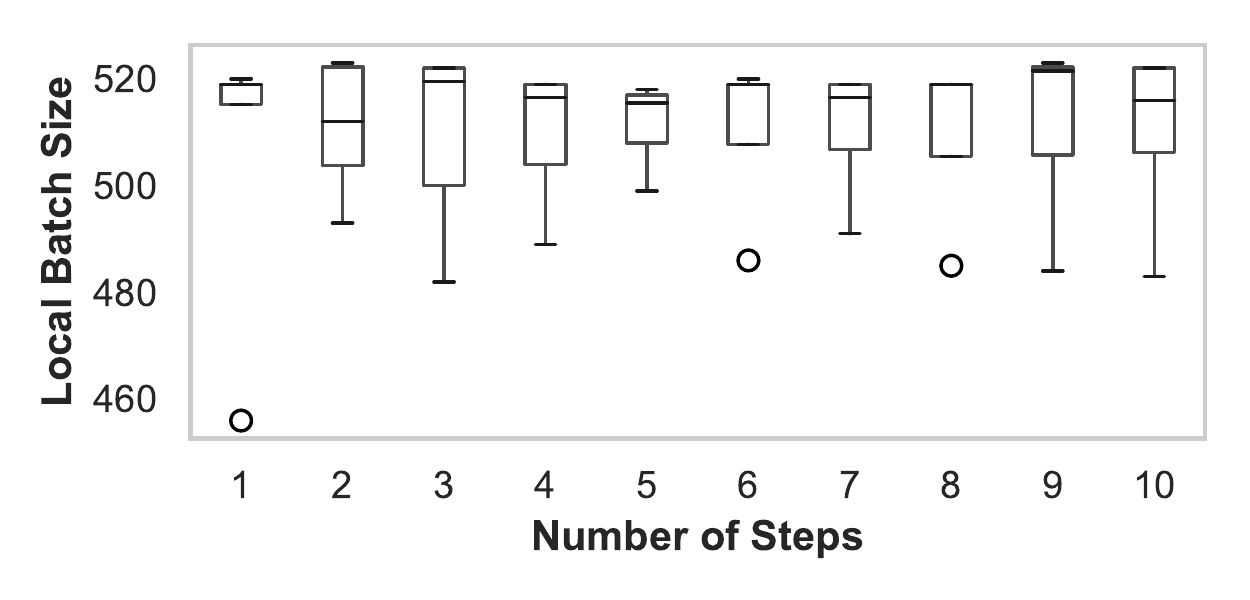}
    \caption{Distributions of per-GPU batch sizes in different training steps (dots are outliers).}
    \label{fig:batch}
\end{figure}

\textbf{Per-GPU batch sizes after trade-off}
Since in the data load balancing optimization, we trade off computation balance with load balance, the imbalance of training local batch size should be studied to understand how much the training batch size is affected.
Therefore, we measure the number of data samples loaded from the PFS for each process at each step.
The simulation is performed on 16 processes and the local batch size is 512. The standard deviation from the first step to the 10th step ranges from 
7.00 to 16.42.
As shown in Figure~\ref{fig:batch}, the local batch sizes are imbalanced, but the distribution is concentrated around 512 (the original local batch size).
This demonstrates that after data load balancing, the training batch size is not very imbalanced,
meaning that trading off the computation balance for load balance does not significantly affect the computation across GPUs.
\section{Related Work}\label{sec:related}

Since optimizations on CPUs/GPUs and communication enable today's DNN training to achieve unprecedented throughput \cite{zhang2021clicktrain,jin2021comet,jiang2020unified,liu2021accelerating,peng2019generic,yin2021parax}, I/O is left behind and becomes the major bottleneck in DNN training.
Furthermore, distributed training techniques enable further acceleration of computation.
Meanwhile, I/O is unable to keep up with the improvement, and it slows down the whole training process.
Therefore, many works focus on data-loading optimizations for distributed DNN training.
We summarize the differences between existing works and our framework in Table~\ref{table:frameworks}.

\newcommand{\TheCell}[2][11em]{
\begin{minipage}{#1}
#2
\end{minipage}
}

\begin{table}[ht!]
\vspace{-2mm}
\renewcommand{\baselinestretch}{.7}
\setlength\extrarowheight{5pt}
\rowcolors{2}{gray!15}{white}
    \TableStyle
    \caption{Summary of techniques used in different methods.}
    \vspace{-11ex}
    \begin{tabular}{ >{\bfseries}l cccccc }
                                     &
        \bfseries\footnotesize \rotatebox{90}{\TheCell[6em]{PyTorch DataLoader~\cite{paszke2019PyTorch}}} 
                                     &
        \bfseries \rotatebox{90}{\TheCell[6em]{Locality-Aware~\cite{yang2019accelerating}} }
                                     &
        \bfseries \rotatebox{90}{\TheCell{DeepIO~\cite{zhu2018entropy}}}
                                     &
        \bfseries \rotatebox{90}{\TheCell{NoPFS~\cite{dryden2021clairvoyant}}} 
                                     &
        \bfseries \rotatebox{90}{\TheCell{Lobster~\cite{liu2022lobster}}}
                                        &
        \bfseries \rotatebox{90}{\TheCell{\TheWork{} (Ours)}}
        \\
        \midrule
        \TheCell{Full Random\-i\-za\-tion}
        & $\bullet$ & $\bullet$ &           & $\bullet$ & $\bullet$ & $\bullet$ 
        \\[1ex]
        \TheCell{Epoch Order Optim.}
        &           &           &           &           & & $\bullet$ 
        \\[1ex]
        \TheCell{Node-Sample Mapping}
        &           & $\circ$
                                     & $\circ$                                                   
                                     & $\circ$  
                                     & $\circ$ 
                                     & $\bullet$                                                 
        \\[1ex]
        \TheCell{Load Balancing}
        &           & $\circ$   & $\bullet$ &   & $\bullet$ & $\bullet$ 
        \\[1ex]
        \TheCell{Aggreg. Chunk Loading}
        &           &           &           &           & & $\bullet$ 
        \\[1ex]
        \bottomrule
    \end{tabular}
\\[1ex]
    
        $\bullet$ technique in use\quad
        $\circ$ data loading achieved via inter-node communication
        \vspace{-3mm}
    \label{table:frameworks}
\end{table}

As mentioned in Section~\ref{sec:datasets}, PyTorch Dataloader \cite{ddp} is a widely used approach for efficient data loading in DNN training with PyTorch, while NoPFS \cite{dryden2021clairvoyant} is a state-of-the-art approach to improve the I/O performance via buffer eviction strategy based on the performance model and a heuristic algorithm.

Yang \textit{et al.}~\cite{yang2019accelerating} propose a locality-aware loading strategy. 
Their optimization takes three steps: (1) Global shuffle during each epoch; (2) Determine the distribution of samples on nodes for data reuse; (3) Inter-node communication to exchange the data samples for data load balancing. 
However, the approach introduces communication overhead to reduce data loading time for each iteration.

Zhu \textit{et al.}~\cite{zhu2018entropy} propose DeepIO that provides a shuffle strategy based on Remote Direct Memory Access (RDMA). DeepIO optimizes the data loading  on two aspects: (1) Local shuffling underlying Remote Direct Memory Access (RDMA) that enables a node to access the data sample buffered on another node; (2) Entropy-aware access order optimization that utilizes cross-entropy to arrange the access order and reduce inter-node communication. However, the optimization does not consider the NUMA effect when accessing data via RDMA~\cite{ren2013design}. Furthermore, their work trades off training accuracy to reduce the data I/O overhead.
We exclude Locality-aware and DeepIO in our comparison, because NoPFS outperforms state-of-the-art frameworks including Locality-aware and DeepIO, as confirmed by our experiments.
Moreover, our evaluation result shows that \TheWork{} achieves higher speedup than NoPFS.

Liu \textit{et al.}~\cite{liu2022lobster} propose an approach called Lobster to optimize data loading by (1) dynamically assigning threads to read and pre-process training data to achieve load balance and (2) approximately arranging the data eviction when the buffer is populated.
Regarding data-loading optimization, our approach is orthogonal to Lobster.
Specifically, our approach adjusts the access order and achieves better data reuse, while Lobster adjusts the number of threads assigned for data reading and preprocessing to improve the data loading throughput.
Thus, \TheWork{} and Lobster can be combined to further improve the end-to-end I/O performance for DNN training. 
Furthermore, for data-load balance, \TheWork{} trades computation balance, while Lobster trades thread balance. However, \TheWork{}'s trade-off strategy is better than Lobster when data loading takes up most of the training time (e.g., 80\%) because in Lobster the number of threads is imbalanced across multiple GPUs most of the time.


Luan \textit{et al.}~\cite{luan2022exoshuffle} propose Exoshuffle, a shuffle system for the MapReduce framework \cite{condie2010mapreduce} that decouples the shuffle control and data planes.
Exoshuffle can be used to shuffle the training data stored on each node and reduce training latency by transferring the data entirely in memory and overlapping shuffling with computation. However, when the size of the training data is larger than the aggregated memory size, the training data must be kept in a remote storage system such as the PFS. Thus, Exoshuffle (in-memory shuffling) does not work in this scenario. 
Moreover, it does not consider data locality to reduce inter-node data movements. 


The NVIDIA Data Loading Library (DALI)~\cite{NVIDIADe15:DALI} is a data loading library that improves the data preprocessing throughput by moving the task from CPU to GPU.
We do not compare with DALI because DALI is orthogonal to \TheWork{} and can be applied to our work to further improve training performance.

\section{Conclusion and Future Work}
\label{sec:conclusion}

CNN-based surrogate models are gaining more and more attention because they can provide approximation results to replace traditional expensive numerical algorithms.
Motivated by the benchmark results showing that data loading is the major performance bottleneck for distributed training of surrogate models, we design \TheWork{}, which consists of offline scheduling and runtime buffer strategies, to optimize the data loading performance by improving the data reuse rate and loading the data in chunks.
Experimental evaluation illustrates that \TheWork{} achieves up to 24.4$\times$ and 3.3$\times$ speedup over a widely used industry framework and a state-of-the-art approach, respectively.
\TheWork{} is implemented on PyTorch DataLoader and H5py (an HDF5 python interface), which makes it easy to use with minimal changes to existing scripts.

\TheWork{} has two limitations: (1) It introduces a one-time offline scheduling overhead, while this computation cost can be shared by multiple models trained on the same dataset. 
Moreover, the scheduling stage can be overlapped with the first epoch, since the dataset is initially stored on the PFS and the data loading time on the first epoch still dominates the training time. (2) It does not implement a multi-layer buffering strategy for HPC hierarchical storage architectures. While our custom buffer is compatible with buffering the data on different node-local storage layers. 

Future work includes but is not limited to the following: 
(1) implement our approach in combination with Lobster and DALI to further improve the performance of data pre-processing and CPU-GPU data loading, respectively;
(2) optimize the offline and runtime scheduling overhead to further improve the performance, then provide the framework as a python interface to improve the performance of existing distributed training works.

\section*{Acknowledgments}

\small This research was supported by the U.S. Department of Energy, Office of Science, Advanced Scientific Computing Research (ASCR), and Office of Science User Facilities, Office of Basic Energy Sciences, under Contract DE-AC02-06CH11357. This work was also supported by the National Science Foundation under Grant 2303820. We gratefully acknowledge the computing resources provided by Argonne Leadership Computing Facility. 

\newpage
\bibliographystyle{ACM-Reference-Format}
\bibliography{refs}


\begin{thebibliography}{51}


\ifx \showCODEN    \undefined \def \showCODEN     #1{\unskip}     \fi
\ifx \showDOI      \undefined \def \showDOI       #1{#1}\fi
\ifx \showISBNx    \undefined \def \showISBNx     #1{\unskip}     \fi
\ifx \showISBNxiii \undefined \def \showISBNxiii  #1{\unskip}     \fi
\ifx \showISSN     \undefined \def \showISSN      #1{\unskip}     \fi
\ifx \showLCCN     \undefined \def \showLCCN      #1{\unskip}     \fi
\ifx \shownote     \undefined \def \shownote      #1{#1}          \fi
\ifx \showarticletitle \undefined \def \showarticletitle #1{#1}   \fi
\ifx \showURL      \undefined \def \showURL       {\relax}        \fi
\providecommand\bibfield[2]{#2}
\providecommand\bibinfo[2]{#2}
\providecommand\natexlab[1]{#1}
\providecommand\showeprint[2][]{arXiv:#2}

\bibitem[Pty({[n.\,d.]})]%
        {PtychoNN}
 \bibinfo{year}{[n.\,d.]}\natexlab{}.
\newblock \bibinfo{title}{{PtychoNN}: Deep learning of ptychographic imaging}.
\newblock
\newblock
\urldef\tempurl%
\url{https://github.com/mcherukara/PtychoNN}
\showURL{%
\tempurl}


\bibitem[the({[n.\,d.]})]%
        {thetaGPU}
 \bibinfo{year}{[n.\,d.]}\natexlab{}.
\newblock \bibinfo{title}{Theta -- Argonne Leadership Computing Facility}.
\newblock
\newblock
\urldef\tempurl%
\url{https://www.alcf.anl.gov/support-center/theta/theta-thetagpu-overview}
\showURL{%
\tempurl}


\bibitem[Abadi et~al\mbox{.}(2016)]%
        {abadi2016TensorFlow}
\bibfield{author}{\bibinfo{person}{Mart{\'\i}n Abadi}, \bibinfo{person}{Paul
  Barham}, \bibinfo{person}{Jianmin Chen}, \bibinfo{person}{Zhifeng Chen},
  \bibinfo{person}{Andy Davis}, \bibinfo{person}{Jeffrey Dean},
  \bibinfo{person}{Matthieu Devin}, \bibinfo{person}{Sanjay Ghemawat},
  \bibinfo{person}{Geoffrey Irving}, \bibinfo{person}{Michael Isard},
  {et~al\mbox{.}}} \bibinfo{year}{2016}\natexlab{}.
\newblock \showarticletitle{{TensorFlow}: A system for large-scale machine
  learning}. In \bibinfo{booktitle}{\emph{12th USENIX symposium on operating
  systems design and implementation (OSDI 16)}}. \bibinfo{pages}{265--283}.
\newblock


\bibitem[Alshemali and Kalita(2020)]%
        {alshemali2020improving}
\bibfield{author}{\bibinfo{person}{Basemah Alshemali} {and}
  \bibinfo{person}{Jugal Kalita}.} \bibinfo{year}{2020}\natexlab{}.
\newblock \showarticletitle{Improving the reliability of deep neural networks
  in {NLP}: A review}.
\newblock \bibinfo{journal}{\emph{Knowledge-Based Systems}}
  \bibinfo{volume}{191} (\bibinfo{year}{2020}), \bibinfo{pages}{105210}.
\newblock


\bibitem[Byna et~al\mbox{.}(2020)]%
        {byna2020exahdf5}
\bibfield{author}{\bibinfo{person}{Suren Byna}, \bibinfo{person}{M~Scot
  Breitenfeld}, \bibinfo{person}{Bin Dong}, \bibinfo{person}{Quincey Koziol},
  \bibinfo{person}{Elena Pourmal}, \bibinfo{person}{Dana Robinson},
  \bibinfo{person}{Jerome Soumagne}, \bibinfo{person}{Houjun Tang},
  \bibinfo{person}{Venkatram Vishwanath}, {and} \bibinfo{person}{Richard
  Warren}.} \bibinfo{year}{2020}\natexlab{}.
\newblock \showarticletitle{{ExaHDF5}: Delivering efficient parallel {I/O} on
  exascale computing systems}.
\newblock \bibinfo{journal}{\emph{Journal of Computer Science and Technology}}
  \bibinfo{volume}{35}, \bibinfo{number}{1} (\bibinfo{year}{2020}),
  \bibinfo{pages}{145--160}.
\newblock


\bibitem[Cherukara et~al\mbox{.}(2020)]%
        {cherukara2020ai}
\bibfield{author}{\bibinfo{person}{Mathew~J Cherukara}, \bibinfo{person}{Tao
  Zhou}, \bibinfo{person}{Youssef Nashed}, \bibinfo{person}{Pablo Enfedaque},
  \bibinfo{person}{Alex Hexemer}, \bibinfo{person}{Ross~J Harder}, {and}
  \bibinfo{person}{Martin~V Holt}.} \bibinfo{year}{2020}\natexlab{}.
\newblock \showarticletitle{{AI}-enabled high-resolution scanning coherent
  diffraction imaging}.
\newblock \bibinfo{journal}{\emph{Applied Physics Letters}}
  \bibinfo{volume}{117}, \bibinfo{number}{4} (\bibinfo{year}{2020}),
  \bibinfo{pages}{044103}.
\newblock


\bibitem[Collins and Duffy(2001)]%
        {collins2001convolution}
\bibfield{author}{\bibinfo{person}{Michael Collins} {and}
  \bibinfo{person}{Nigel Duffy}.} \bibinfo{year}{2001}\natexlab{}.
\newblock \showarticletitle{Convolution kernels for natural language}.
\newblock \bibinfo{journal}{\emph{Advances in neural information processing
  systems}}  \bibinfo{volume}{14} (\bibinfo{year}{2001}).
\newblock


\bibitem[Condie et~al\mbox{.}(2010)]%
        {condie2010mapreduce}
\bibfield{author}{\bibinfo{person}{Tyson Condie}, \bibinfo{person}{Neil
  Conway}, \bibinfo{person}{Peter Alvaro}, \bibinfo{person}{Joseph~M
  Hellerstein}, \bibinfo{person}{Khaled Elmeleegy}, {and}
  \bibinfo{person}{Russell Sears}.} \bibinfo{year}{2010}\natexlab{}.
\newblock \showarticletitle{MapReduce online}. In
  \bibinfo{booktitle}{\emph{NSDI}}, Vol.~\bibinfo{volume}{10}.
  \bibinfo{pages}{20}.
\newblock


\bibitem[Deng et~al\mbox{.}(2009)]%
        {deng2009imagenet}
\bibfield{author}{\bibinfo{person}{Jia Deng}, \bibinfo{person}{Wei Dong},
  \bibinfo{person}{Richard Socher}, \bibinfo{person}{Li-Jia Li},
  \bibinfo{person}{Kai Li}, {and} \bibinfo{person}{Li Fei-Fei}.}
  \bibinfo{year}{2009}\natexlab{}.
\newblock \showarticletitle{{ImageNet}: A large-scale hierarchical image
  database}. In \bibinfo{booktitle}{\emph{2009 IEEE conference on computer
  vision and pattern recognition}}. IEEE, \bibinfo{pages}{248--255}.
\newblock


\bibitem[Dong et~al\mbox{.}(2019)]%
        {dong2019adaptive}
\bibfield{author}{\bibinfo{person}{Wenqian Dong}, \bibinfo{person}{Jie Liu},
  \bibinfo{person}{Zhen Xie}, {and} \bibinfo{person}{Dong Li}.}
  \bibinfo{year}{2019}\natexlab{}.
\newblock \showarticletitle{Adaptive neural network-based approximation to
  accelerate {Eulerian} fluid simulation}. In
  \bibinfo{booktitle}{\emph{Proceedings of the International Conference for
  High Performance Computing, Networking, Storage and Analysis}}.
  \bibinfo{pages}{1--22}.
\newblock


\bibitem[Dong et~al\mbox{.}(2020)]%
        {dong2020smart}
\bibfield{author}{\bibinfo{person}{Wenqian Dong}, \bibinfo{person}{Zhen Xie},
  \bibinfo{person}{Gokcen Kestor}, {and} \bibinfo{person}{Dong Li}.}
  \bibinfo{year}{2020}\natexlab{}.
\newblock \showarticletitle{Smart-PGSim: Using neural network to accelerate
  {AC-OPF} power grid simulation}. In \bibinfo{booktitle}{\emph{SC20:
  International Conference for High Performance Computing, Networking, Storage
  and Analysis}}. IEEE, \bibinfo{pages}{1--15}.
\newblock


\bibitem[Dryden et~al\mbox{.}(2021)]%
        {dryden2021clairvoyant}
\bibfield{author}{\bibinfo{person}{Nikoli Dryden}, \bibinfo{person}{Roman
  B{\"o}hringer}, \bibinfo{person}{Tal Ben-Nun}, {and} \bibinfo{person}{Torsten
  Hoefler}.} \bibinfo{year}{2021}\natexlab{}.
\newblock \showarticletitle{Clairvoyant prefetching for distributed machine
  learning I/O}. In \bibinfo{booktitle}{\emph{Proceedings of the International
  Conference for High Performance Computing, Networking, Storage and
  Analysis}}. \bibinfo{pages}{1--15}.
\newblock


\bibitem[Farrell et~al\mbox{.}(2021)]%
        {farrell2021mlperf}
\bibfield{author}{\bibinfo{person}{Steven Farrell}, \bibinfo{person}{Murali
  Emani}, \bibinfo{person}{Jacob Balma}, \bibinfo{person}{Lukas Drescher},
  \bibinfo{person}{Aleksandr Drozd}, \bibinfo{person}{Andreas Fink},
  \bibinfo{person}{Geoffrey Fox}, \bibinfo{person}{David Kanter},
  \bibinfo{person}{Thorsten Kurth}, \bibinfo{person}{Peter Mattson},
  {et~al\mbox{.}}} \bibinfo{year}{2021}\natexlab{}.
\newblock \showarticletitle{MLPerf™ HPC: A Holistic Benchmark Suite for
  Scientific Machine Learning on HPC Systems}. In
  \bibinfo{booktitle}{\emph{2021 IEEE/ACM Workshop on Machine Learning in High
  Performance Computing Environments (MLHPC)}}. IEEE, \bibinfo{pages}{33--45}.
\newblock


\bibitem[Flood(1956)]%
        {flood1956traveling}
\bibfield{author}{\bibinfo{person}{Merrill~M Flood}.}
  \bibinfo{year}{1956}\natexlab{}.
\newblock \showarticletitle{The traveling-salesman problem}.
\newblock \bibinfo{journal}{\emph{Operations research}} \bibinfo{volume}{4},
  \bibinfo{number}{1} (\bibinfo{year}{1956}), \bibinfo{pages}{61--75}.
\newblock


\bibitem[He et~al\mbox{.}(2016)]%
        {he2016deep}
\bibfield{author}{\bibinfo{person}{Kaiming He}, \bibinfo{person}{Xiangyu
  Zhang}, \bibinfo{person}{Shaoqing Ren}, {and} \bibinfo{person}{Jian Sun}.}
  \bibinfo{year}{2016}\natexlab{}.
\newblock \showarticletitle{Deep residual learning for image recognition}. In
  \bibinfo{booktitle}{\emph{Proceedings of the IEEE/CVF Conference on Computer
  Vision and Pattern Recognition}}. \bibinfo{pages}{770--778}.
\newblock


\bibitem[He et~al\mbox{.}(2020)]%
        {he2020lightgcn}
\bibfield{author}{\bibinfo{person}{Xiangnan He}, \bibinfo{person}{Kuan Deng},
  \bibinfo{person}{Xiang Wang}, \bibinfo{person}{Yan Li},
  \bibinfo{person}{Yongdong Zhang}, {and} \bibinfo{person}{Meng Wang}.}
  \bibinfo{year}{2020}\natexlab{}.
\newblock \showarticletitle{{LightGCN}: Simplifying and powering graph
  convolution network for recommendation}. In
  \bibinfo{booktitle}{\emph{Proceedings of the 43rd International ACM SIGIR
  conference on research and development in Information Retrieval}}.
  \bibinfo{pages}{639--648}.
\newblock


\bibitem[Jacobs et~al\mbox{.}(2019)]%
        {jacobs2019parallelizing}
\bibfield{author}{\bibinfo{person}{Sam~Ade Jacobs}, \bibinfo{person}{Brian
  Van~Essen}, \bibinfo{person}{David Hysom}, \bibinfo{person}{Jae-Seung Yeom},
  \bibinfo{person}{Tim Moon}, \bibinfo{person}{Rushil Anirudh},
  \bibinfo{person}{Jayaraman~J Thiagaranjan}, \bibinfo{person}{Shusen Liu},
  \bibinfo{person}{Peer-Timo Bremer}, \bibinfo{person}{Jim Gaffney},
  {et~al\mbox{.}}} \bibinfo{year}{2019}\natexlab{}.
\newblock \showarticletitle{Parallelizing training of deep generative models on
  massive scientific datasets}. In \bibinfo{booktitle}{\emph{2019 IEEE
  International Conference on Cluster Computing (CLUSTER)}}. IEEE,
  \bibinfo{pages}{1--10}.
\newblock


\bibitem[Jiang et~al\mbox{.}(2020)]%
        {jiang2020unified}
\bibfield{author}{\bibinfo{person}{Yimin Jiang}, \bibinfo{person}{Yibo Zhu},
  \bibinfo{person}{Chang Lan}, \bibinfo{person}{Bairen Yi},
  \bibinfo{person}{Yong Cui}, {and} \bibinfo{person}{Chuanxiong Guo}.}
  \bibinfo{year}{2020}\natexlab{}.
\newblock \showarticletitle{A unified architecture for accelerating distributed
  {DNN} training in heterogeneous {GPU/CPU} clusters}. In
  \bibinfo{booktitle}{\emph{14th USENIX Symposium on Operating Systems Design
  and Implementation (OSDI 20)}}. \bibinfo{pages}{463--479}.
\newblock


\bibitem[Jin et~al\mbox{.}(2021)]%
        {jin2021comet}
\bibfield{author}{\bibinfo{person}{Sian Jin}, \bibinfo{person}{Chengming
  Zhang}, \bibinfo{person}{Xintong Jiang}, \bibinfo{person}{Yunhe Feng},
  \bibinfo{person}{Hui Guan}, \bibinfo{person}{Guanpeng Li},
  \bibinfo{person}{Shuaiwen~Leon Song}, {and} \bibinfo{person}{Dingwen Tao}.}
  \bibinfo{year}{2021}\natexlab{}.
\newblock \showarticletitle{COMET: A novel memory-efficient deep learning
  training framework by using error-bounded lossy compression}.
\newblock \bibinfo{journal}{\emph{Proceedings of the VLDB Endowment}}
  \bibinfo{volume}{15}, \bibinfo{number}{4} (\bibinfo{year}{2021}),
  \bibinfo{pages}{886--899}.
\newblock


\bibitem[Kennedy and Eberhart(1995)]%
        {pso}
\bibfield{author}{\bibinfo{person}{James Kennedy} {and}
  \bibinfo{person}{Russell Eberhart}.} \bibinfo{year}{1995}\natexlab{}.
\newblock \showarticletitle{Particle swarm optimization}. In
  \bibinfo{booktitle}{\emph{Proceedings of ICNN'95-international conference on
  neural networks}}, Vol.~\bibinfo{volume}{4}. IEEE,
  \bibinfo{pages}{1942--1948}.
\newblock


\bibitem[Kurth et~al\mbox{.}(2018)]%
        {kurth2018exascale}
\bibfield{author}{\bibinfo{person}{Thorsten Kurth}, \bibinfo{person}{Sean
  Treichler}, \bibinfo{person}{Joshua Romero}, \bibinfo{person}{Mayur
  Mudigonda}, \bibinfo{person}{Nathan Luehr}, \bibinfo{person}{Everett
  Phillips}, \bibinfo{person}{Ankur Mahesh}, \bibinfo{person}{Michael
  Matheson}, \bibinfo{person}{Jack Deslippe}, \bibinfo{person}{Massimiliano
  Fatica}, {et~al\mbox{.}}} \bibinfo{year}{2018}\natexlab{}.
\newblock \showarticletitle{Exascale deep learning for climate analytics}. In
  \bibinfo{booktitle}{\emph{SC18: International Conference for High Performance
  Computing, Networking, Storage and Analysis}}. IEEE,
  \bibinfo{pages}{649--660}.
\newblock


\bibitem[Li et~al\mbox{.}(2020)]%
        {ddp}
\bibfield{author}{\bibinfo{person}{Shen Li}, \bibinfo{person}{Yanli Zhao},
  \bibinfo{person}{Rohan Varma}, \bibinfo{person}{Omkar Salpekar},
  \bibinfo{person}{Pieter Noordhuis}, \bibinfo{person}{Teng Li},
  \bibinfo{person}{Adam Paszke}, \bibinfo{person}{Jeff Smith},
  \bibinfo{person}{Brian Vaughan}, \bibinfo{person}{Pritam Damania},
  {et~al\mbox{.}}} \bibinfo{year}{2020}\natexlab{}.
\newblock \showarticletitle{PyTorch distributed: experiences on accelerating
  data parallel training}.
\newblock \bibinfo{journal}{\emph{Proceedings of the VLDB Endowment}}
  \bibinfo{volume}{13}, \bibinfo{number}{12} (\bibinfo{year}{2020}),
  \bibinfo{pages}{3005--3018}.
\newblock


\bibitem[Liu et~al\mbox{.}(2021)]%
        {liu2021accelerating}
\bibfield{author}{\bibinfo{person}{Hongyuan Liu}, \bibinfo{person}{Bogdan
  Nicolae}, \bibinfo{person}{Sheng Di}, \bibinfo{person}{Franck Cappello},
  {and} \bibinfo{person}{Adwait Jog}.} \bibinfo{year}{2021}\natexlab{}.
\newblock \showarticletitle{Accelerating DNN architecture search at scale using
  selective weight transfer}. In \bibinfo{booktitle}{\emph{2021 IEEE
  International Conference on Cluster Computing (CLUSTER)}}. IEEE,
  \bibinfo{pages}{82--93}.
\newblock


\bibitem[Liu et~al\mbox{.}(2022)]%
        {liu2022lobster}
\bibfield{author}{\bibinfo{person}{Jie Liu}, \bibinfo{person}{Bogdan Nicolae},
  {and} \bibinfo{person}{Dong Li}.} \bibinfo{year}{2022}\natexlab{}.
\newblock \showarticletitle{Lobster: Load Balance-Aware I/O for Distributed DNN
  Training}. In \bibinfo{booktitle}{\emph{ICPP'22: The 51st International
  Conference on Parallel Processing}}.
\newblock


\bibitem[Luan et~al\mbox{.}(2022)]%
        {luan2022exoshuffle}
\bibfield{author}{\bibinfo{person}{Frank~Sifei Luan},
  \bibinfo{person}{Stephanie Wang}, \bibinfo{person}{Samyukta Yagati},
  \bibinfo{person}{Sean Kim}, \bibinfo{person}{Kenneth Lien},
  \bibinfo{person}{SangBin Cho}, \bibinfo{person}{Eric Liang}, {and}
  \bibinfo{person}{Ion Stoica}.} \bibinfo{year}{2022}\natexlab{}.
\newblock \showarticletitle{Exoshuffle: Large-Scale Shuffle at the Application
  Level}.
\newblock \bibinfo{journal}{\emph{arXiv preprint arXiv:2203.05072}}
  (\bibinfo{year}{2022}).
\newblock


\bibitem[Mathuriya et~al\mbox{.}(2018)]%
        {mathuriya2018cosmoflow}
\bibfield{author}{\bibinfo{person}{Amrita Mathuriya}, \bibinfo{person}{Deborah
  Bard}, \bibinfo{person}{Peter Mendygral}, \bibinfo{person}{Lawrence Meadows},
  \bibinfo{person}{James Arnemann}, \bibinfo{person}{Lei Shao},
  \bibinfo{person}{Siyu He}, \bibinfo{person}{Tuomas K{\"a}rn{\"a}},
  \bibinfo{person}{Diana Moise}, \bibinfo{person}{Simon~J Pennycook},
  {et~al\mbox{.}}} \bibinfo{year}{2018}\natexlab{}.
\newblock \showarticletitle{CosmoFlow: Using deep learning to learn the
  universe at scale}. In \bibinfo{booktitle}{\emph{SC18: International
  Conference for High Performance Computing, Networking, Storage and
  Analysis}}. IEEE, \bibinfo{pages}{819--829}.
\newblock


\bibitem[Noirot et~al\mbox{.}(2018)]%
        {noirot2018workshop}
\bibfield{author}{\bibinfo{person}{Philippe Noirot}, \bibinfo{person}{Andrzej
  Joachimiak}, {and} \bibinfo{person}{Robert~F Fischetti}.}
  \bibinfo{year}{2018}\natexlab{}.
\newblock \bibinfo{booktitle}{\emph{Workshop on Biological Science
  Opportunities Provided by the APS Upgrade}}.
\newblock \bibinfo{type}{{T}echnical {R}eport}. \bibinfo{institution}{Argonne
  National Laboratory (ANL), Lemont, IL (United States)}.
\newblock


\bibitem[NVIDIA({[n.\,d.]})]%
        {NVIDIADe15:DALI}
\bibfield{author}{\bibinfo{person}{NVIDIA}.}
  \bibinfo{year}{[n.\,d.]}\natexlab{}.
\newblock \bibinfo{title}{NVIDIA Developer Data Loading Library (DALI)}.
\newblock
  \bibinfo{howpublished}{\url{https://docs.nvidia.com/deeplearning/dali/user-guide/docs/}}.
\newblock
\newblock
\shownote{(Accessed on 10/27/2022)}.


\bibitem[Obiols-Sales et~al\mbox{.}(2020)]%
        {obiols2020cfdnet}
\bibfield{author}{\bibinfo{person}{Octavi Obiols-Sales},
  \bibinfo{person}{Abhinav Vishnu}, \bibinfo{person}{Nicholas Malaya}, {and}
  \bibinfo{person}{Aparna Chandramowliswharan}.}
  \bibinfo{year}{2020}\natexlab{}.
\newblock \showarticletitle{CFDNet: A deep learning-based accelerator for fluid
  simulations}. In \bibinfo{booktitle}{\emph{Proceedings of the 34th ACM
  international conference on supercomputing}}. \bibinfo{pages}{1--12}.
\newblock


\bibitem[Oyama et~al\mbox{.}(2020)]%
        {oyama2020case}
\bibfield{author}{\bibinfo{person}{Yosuke Oyama}, \bibinfo{person}{Naoya
  Maruyama}, \bibinfo{person}{Nikoli Dryden}, \bibinfo{person}{Erin McCarthy},
  \bibinfo{person}{Peter Harrington}, \bibinfo{person}{Jan Balewski},
  \bibinfo{person}{Satoshi Matsuoka}, \bibinfo{person}{Peter Nugent}, {and}
  \bibinfo{person}{Brian Van~Essen}.} \bibinfo{year}{2020}\natexlab{}.
\newblock \showarticletitle{The case for strong scaling in deep learning:
  Training large {3D} {CNNs} with hybrid parallelism}.
\newblock \bibinfo{journal}{\emph{IEEE Transactions on Parallel and Distributed
  Systems}} \bibinfo{volume}{32}, \bibinfo{number}{7} (\bibinfo{year}{2020}),
  \bibinfo{pages}{1641--1652}.
\newblock


\bibitem[Papadimitriou(1977)]%
        {TSP_npc}
\bibfield{author}{\bibinfo{person}{Christos~H Papadimitriou}.}
  \bibinfo{year}{1977}\natexlab{}.
\newblock \showarticletitle{The Euclidean travelling salesman problem is
  NP-complete}.
\newblock \bibinfo{journal}{\emph{Theoretical computer science}}
  \bibinfo{volume}{4}, \bibinfo{number}{3} (\bibinfo{year}{1977}),
  \bibinfo{pages}{237--244}.
\newblock


\bibitem[Paszke et~al\mbox{.}(2019)]%
        {paszke2019PyTorch}
\bibfield{author}{\bibinfo{person}{Adam Paszke}, \bibinfo{person}{Sam Gross},
  \bibinfo{person}{Francisco Massa}, \bibinfo{person}{Adam Lerer},
  \bibinfo{person}{James Bradbury}, \bibinfo{person}{Gregory Chanan},
  \bibinfo{person}{Trevor Killeen}, \bibinfo{person}{Zeming Lin},
  \bibinfo{person}{Natalia Gimelshein}, \bibinfo{person}{Luca Antiga},
  {et~al\mbox{.}}} \bibinfo{year}{2019}\natexlab{}.
\newblock \showarticletitle{{PyTorch}: An imperative style, high-performance
  deep learning library}.
\newblock \bibinfo{journal}{\emph{Advances in neural information processing
  systems}}  \bibinfo{volume}{32} (\bibinfo{year}{2019}).
\newblock


\bibitem[Peng et~al\mbox{.}(2019)]%
        {peng2019generic}
\bibfield{author}{\bibinfo{person}{Yanghua Peng}, \bibinfo{person}{Yibo Zhu},
  \bibinfo{person}{Yangrui Chen}, \bibinfo{person}{Yixin Bao},
  \bibinfo{person}{Bairen Yi}, \bibinfo{person}{Chang Lan},
  \bibinfo{person}{Chuan Wu}, {and} \bibinfo{person}{Chuanxiong Guo}.}
  \bibinfo{year}{2019}\natexlab{}.
\newblock \showarticletitle{A generic communication scheduler for distributed
  {DNN} training acceleration}. In \bibinfo{booktitle}{\emph{Proceedings of the
  27th ACM Symposium on Operating Systems Principles}}.
  \bibinfo{pages}{16--29}.
\newblock


\bibitem[Pumma et~al\mbox{.}(2017)]%
        {pumma2017towards}
\bibfield{author}{\bibinfo{person}{Sarunya Pumma}, \bibinfo{person}{Min Si},
  \bibinfo{person}{Wu-chun Feng}, {and} \bibinfo{person}{Pavan Balaji}.}
  \bibinfo{year}{2017}\natexlab{}.
\newblock \showarticletitle{Towards scalable deep learning via I/O analysis and
  optimization}. In \bibinfo{booktitle}{\emph{19th IEEE International
  Conference on High Performance Computing and Communications; 15th IEEE
  International Conference on Smart City; 3rd IEEE International Conference on
  Data Science and Systems (HPCC/SmartCity/DSS)}}. IEEE,
  \bibinfo{pages}{223--230}.
\newblock


\bibitem[Pumma et~al\mbox{.}(2019)]%
        {pumma2019scalable}
\bibfield{author}{\bibinfo{person}{Sarunya Pumma}, \bibinfo{person}{Min Si},
  \bibinfo{person}{Wu-Chun Feng}, {and} \bibinfo{person}{Pavan Balaji}.}
  \bibinfo{year}{2019}\natexlab{}.
\newblock \showarticletitle{Scalable deep learning via I/O analysis and
  optimization}.
\newblock \bibinfo{journal}{\emph{ACM Transactions on Parallel Computing
  (TOPC)}} \bibinfo{volume}{6}, \bibinfo{number}{2} (\bibinfo{year}{2019}),
  \bibinfo{pages}{1--34}.
\newblock


\bibitem[Ren et~al\mbox{.}(2013)]%
        {ren2013design}
\bibfield{author}{\bibinfo{person}{Yufei Ren}, \bibinfo{person}{Tan Li},
  \bibinfo{person}{Dantong Yu}, \bibinfo{person}{Shudong Jin}, {and}
  \bibinfo{person}{Thomas Robertazzi}.} \bibinfo{year}{2013}\natexlab{}.
\newblock \showarticletitle{Design and performance evaluation of NUMA-aware
  RDMA-based end-to-end data transfer systems}. In
  \bibinfo{booktitle}{\emph{Proceedings of the International Conference on High
  Performance Computing, Networking, Storage and Analysis}}.
  \bibinfo{pages}{1--10}.
\newblock


\bibitem[Sergeev and Del~Balso(2018)]%
        {sergeev2018horovod}
\bibfield{author}{\bibinfo{person}{Alexander Sergeev} {and}
  \bibinfo{person}{Mike Del~Balso}.} \bibinfo{year}{2018}\natexlab{}.
\newblock \showarticletitle{Horovod: Fast and easy distributed deep learning in
  {TensorFlow}}.
\newblock \bibinfo{journal}{\emph{arXiv preprint arXiv:1802.05799}}
  (\bibinfo{year}{2018}).
\newblock


\bibitem[Shi et~al\mbox{.}(2022)]%
        {shi2022gnn}
\bibfield{author}{\bibinfo{person}{Neng Shi}, \bibinfo{person}{Jiayi Xu},
  \bibinfo{person}{Skylar~W Wurster}, \bibinfo{person}{Hanqi Guo},
  \bibinfo{person}{Jonathan Woodring}, \bibinfo{person}{Luke~P Van~Roekel},
  {and} \bibinfo{person}{Han-Wei Shen}.} \bibinfo{year}{2022}\natexlab{}.
\newblock \showarticletitle{GNN-Surrogate: A Hierarchical and Adaptive Graph
  Neural Network for Parameter Space Exploration of Unstructured-Mesh Ocean
  Simulations}.
\newblock \bibinfo{journal}{\emph{IEEE Transactions on Visualization and
  Computer Graphics}} \bibinfo{volume}{28}, \bibinfo{number}{6}
  (\bibinfo{year}{2022}), \bibinfo{pages}{2301--2313}.
\newblock


\bibitem[Shi et~al\mbox{.}(2007)]%
        {shi2007particle}
\bibfield{author}{\bibinfo{person}{Xiaohu~H Shi}, \bibinfo{person}{Yanchun~Chun
  Liang}, \bibinfo{person}{Heow~Pueh Lee}, \bibinfo{person}{C Lu}, {and}
  \bibinfo{person}{QX Wang}.} \bibinfo{year}{2007}\natexlab{}.
\newblock \showarticletitle{Particle swarm optimization-based algorithms for
  TSP and generalized TSP}.
\newblock \bibinfo{journal}{\emph{Information processing letters}}
  \bibinfo{volume}{103}, \bibinfo{number}{5} (\bibinfo{year}{2007}),
  \bibinfo{pages}{169--176}.
\newblock


\bibitem[{The HDF Group}(2010)]%
        {hdf5}
\bibfield{author}{\bibinfo{person}{{The HDF Group}}.}
  \bibinfo{year}{2000-2010}\natexlab{}.
\newblock \bibinfo{title}{Hierarchical Data Format version 5}.
\newblock
\newblock
\urldef\tempurl%
\url{http://www.hdfgroup.org/HDF5}
\showURL{%
\tempurl}


\bibitem[{UIUC}({[n.\,d.]})]%
        {recorder}
\bibfield{author}{\bibinfo{person}{{UIUC}}.}
  \bibinfo{year}{[n.\,d.]}\natexlab{}.
\newblock \bibinfo{title}{Multi-level tracing library}.
\newblock \bibinfo{howpublished}{\url{https://github.com/uiuc-hpc/Recorder}}.
\newblock


\bibitem[Voulodimos et~al\mbox{.}(2018)]%
        {voulodimos2018deep}
\bibfield{author}{\bibinfo{person}{Athanasios Voulodimos},
  \bibinfo{person}{Nikolaos Doulamis}, \bibinfo{person}{Anastasios Doulamis},
  {and} \bibinfo{person}{Eftychios Protopapadakis}.}
  \bibinfo{year}{2018}\natexlab{}.
\newblock \showarticletitle{Deep learning for computer vision: A brief review}.
\newblock \bibinfo{journal}{\emph{Computational intelligence and neuroscience}}
   \bibinfo{volume}{2018} (\bibinfo{year}{2018}).
\newblock


\bibitem[Yang and Cong(2019)]%
        {yang2019accelerating}
\bibfield{author}{\bibinfo{person}{Chih-Chieh Yang} {and}
  \bibinfo{person}{Guojing Cong}.} \bibinfo{year}{2019}\natexlab{}.
\newblock \showarticletitle{Accelerating data loading in deep neural network
  training}. In \bibinfo{booktitle}{\emph{2019 IEEE 26th International
  Conference on High Performance Computing, Data, and Analytics (HiPC)}}. IEEE,
  \bibinfo{pages}{235--245}.
\newblock


\bibitem[Yao et~al\mbox{.}(2022)]%
        {yao2022autophasenn}
\bibfield{author}{\bibinfo{person}{Yudong Yao}, \bibinfo{person}{Henry Chan},
  \bibinfo{person}{Subramanian Sankaranarayanan}, \bibinfo{person}{Prasanna
  Balaprakash}, \bibinfo{person}{Ross~J Harder}, {and}
  \bibinfo{person}{Mathew~J Cherukara}.} \bibinfo{year}{2022}\natexlab{}.
\newblock \showarticletitle{AutoPhaseNN: Unsupervised physics-aware deep
  learning of {3D} nanoscale {Bragg} coherent diffraction imaging}.
\newblock \bibinfo{journal}{\emph{npj Computational Materials}}
  \bibinfo{volume}{8}, \bibinfo{number}{1} (\bibinfo{year}{2022}),
  \bibinfo{pages}{1--8}.
\newblock


\bibitem[Yin et~al\mbox{.}(2022)]%
        {yin2022strategies}
\bibfield{author}{\bibinfo{person}{Junqi Yin}, \bibinfo{person}{Feiyi Wang},
  {and} \bibinfo{person}{Mallikarjun Shankar}.}
  \bibinfo{year}{2022}\natexlab{}.
\newblock \showarticletitle{Strategies for Integrating Deep Learning Surrogate
  Models with HPC Simulation Applications}. In \bibinfo{booktitle}{\emph{2022
  IEEE International Parallel and Distributed Processing Symposium Workshops
  (IPDPSW)}}. IEEE, \bibinfo{pages}{01--10}.
\newblock


\bibitem[Yin et~al\mbox{.}(2021)]%
        {yin2021parax}
\bibfield{author}{\bibinfo{person}{Lujia Yin}, \bibinfo{person}{Yiming Zhang},
  \bibinfo{person}{Zhaoning Zhang}, \bibinfo{person}{Yuxing Peng}, {and}
  \bibinfo{person}{Peng Zhao}.} \bibinfo{year}{2021}\natexlab{}.
\newblock \showarticletitle{ParaX: Boosting deep learning for big data
  analytics on many-core {CPUs}}.
\newblock \bibinfo{journal}{\emph{Proceedings of the VLDB Endowment}}
  \bibinfo{volume}{14}, \bibinfo{number}{6} (\bibinfo{year}{2021}),
  \bibinfo{pages}{864--877}.
\newblock


\bibitem[Yu et~al\mbox{.}(2022)]%
        {yu2022scalable}
\bibfield{author}{\bibinfo{person}{Xiaodong Yu}, \bibinfo{person}{Viktor
  Nikitin}, \bibinfo{person}{Daniel~J Ching}, \bibinfo{person}{Selin Aslan},
  \bibinfo{person}{Do{\u{g}}a G{\"u}rsoy}, {and} \bibinfo{person}{Tekin
  Bi{\c{c}}er}.} \bibinfo{year}{2022}\natexlab{}.
\newblock \showarticletitle{Scalable and accurate multi-GPU-based image
  reconstruction of large-scale ptychography data}.
\newblock \bibinfo{journal}{\emph{Scientific Reports}} \bibinfo{volume}{12},
  \bibinfo{number}{1} (\bibinfo{year}{2022}), \bibinfo{pages}{1--16}.
\newblock


\bibitem[Zhang et~al\mbox{.}(2021b)]%
        {zhang2021clicktrain}
\bibfield{author}{\bibinfo{person}{Chengming Zhang}, \bibinfo{person}{Geng
  Yuan}, \bibinfo{person}{Wei Niu}, \bibinfo{person}{Jiannan Tian},
  \bibinfo{person}{Sian Jin}, \bibinfo{person}{Donglin Zhuang},
  \bibinfo{person}{Zhe Jiang}, \bibinfo{person}{Yanzhi Wang},
  \bibinfo{person}{Bin Ren}, \bibinfo{person}{Shuaiwen~Leon Song},
  {et~al\mbox{.}}} \bibinfo{year}{2021}\natexlab{b}.
\newblock \showarticletitle{Clicktrain: Efficient and accurate end-to-end deep
  learning training via fine-grained architecture-preserving pruning}. In
  \bibinfo{booktitle}{\emph{Proceedings of the ACM International Conference on
  Supercomputing}}. \bibinfo{pages}{266--278}.
\newblock


\bibitem[Zhang et~al\mbox{.}(2019)]%
        {zhang2019deep}
\bibfield{author}{\bibinfo{person}{Shuai Zhang}, \bibinfo{person}{Lina Yao},
  \bibinfo{person}{Aixin Sun}, {and} \bibinfo{person}{Yi Tay}.}
  \bibinfo{year}{2019}\natexlab{}.
\newblock \showarticletitle{Deep learning based recommender system: A survey
  and new perspectives}.
\newblock \bibinfo{journal}{\emph{ACM Computing Surveys (CSUR)}}
  \bibinfo{volume}{52}, \bibinfo{number}{1} (\bibinfo{year}{2019}),
  \bibinfo{pages}{1--38}.
\newblock


\bibitem[Zhang et~al\mbox{.}(2021a)]%
        {zhang2021distributed}
\bibfield{author}{\bibinfo{person}{Yuhao Zhang}, \bibinfo{person}{Frank
  Mcquillan}, \bibinfo{person}{Nandish Jayaram}, \bibinfo{person}{Nikhil Kak},
  \bibinfo{person}{Ekta Khanna}, \bibinfo{person}{Orhan Kislal},
  \bibinfo{person}{Domino Valdano}, {and} \bibinfo{person}{Arun Kumar}.}
  \bibinfo{year}{2021}\natexlab{a}.
\newblock \showarticletitle{Distributed deep learning on data systems: A
  comparative analysis of approaches}.
\newblock \bibinfo{journal}{\emph{Proceedings of the VLDB Endowment}}
  \bibinfo{volume}{14}, \bibinfo{number}{10} (\bibinfo{year}{2021}),
  \bibinfo{pages}{1769--1782}.
\newblock


\bibitem[Zhu et~al\mbox{.}(2018)]%
        {zhu2018entropy}
\bibfield{author}{\bibinfo{person}{Yue Zhu}, \bibinfo{person}{Fahim Chowdhury},
  \bibinfo{person}{Huansong Fu}, \bibinfo{person}{Adam Moody},
  \bibinfo{person}{Kathryn Mohror}, \bibinfo{person}{Kento Sato}, {and}
  \bibinfo{person}{Weikuan Yu}.} \bibinfo{year}{2018}\natexlab{}.
\newblock \showarticletitle{Entropy-aware I/O pipelining for large-scale deep
  learning on HPC systems}. In \bibinfo{booktitle}{\emph{2018 IEEE 26th
  International Symposium on Modeling, Analysis, and Simulation of Computer and
  Telecommunication Systems (MASCOTS)}}. IEEE, \bibinfo{pages}{145--156}.
\newblock


\end{thebibliography}

\end{document}